\documentclass[lettersize,journal]{IEEEtran}
\usepackage{amsmath,amsfonts,amssymb}
\usepackage{algorithmic}
\usepackage{algorithm}
\usepackage{array}
\usepackage{textcomp}
\usepackage{stfloats}
\usepackage{url}
\usepackage{verbatim}
\usepackage{graphicx}
\usepackage{cite}
\usepackage{hyperref}

\usepackage{subcaption}
\usepackage{float}
\usepackage{enumerate}
\usepackage{xcolor}
\usepackage{bm}
\usepackage{tikz}
\usepackage{pgfplots}
\usetikzlibrary{spy, arrows, patterns, arrows.meta}
\pgfdeclarelayer{bg}    
\pgfsetlayers{bg,main}  

\usepackage{tabularray}
\UseTblrLibrary{amsmath,booktabs,counter,diagbox,nameref,siunitx,varwidth,zref}

\newcommand{\nlatentgrids}{L}
\newcommand{\img}{\mathbf{x}}
\newcommand{\codedimg}{\hat{\mathbf{x}}}
\newcommand{\latent}{\mathbf{y}}
\newcommand{\qlatent}{\hat{\latent}}
\newcommand{\denselatent}{\hat{\mathbf{z}}}

\newcommand{\synthparam}{\bm{\theta}}
\newcommand{\armparam}{\bm{\psi}}
\newcommand{\upparam}{\bm{\upsilon}}
\newcommand{\ifceparam}{\bm{\chi}}
\newcommand{\synth}{f_{\synthparam}}
\newcommand{\arm}{f_{\armparam}}
\newcommand{\upsample}{f_{\upparam}}

\definecolor{cc_darkpastelpurple}{rgb}{0.59, 0.44, 0.84}
\definecolor{cc_red}{HTML}{E76F51}
\definecolor{cc_blue}{HTML}{376996}
\definecolor{cc_green}{HTML}{2A9D8F}
\definecolor{cc_purple}{HTML}{822E81}
\definecolor{cc_white}{HTML}{f0dfbb}

\begin{document}

\title{Cool-chic 5.0: Faster Encoding and Inter-Feature Entropy Modeling for Overfitted Image Compression}

\author{Théo Ladune, Pierrick Philippe, Pierre Jaffuer, Théophile Blard,\\ Sylvain Kervadec, Félix Henry, Gordon Clare
\thanks{All authors are with Orange Research, France. Contact email: theo.ladune@orange.com}
}



\maketitle

\begin{abstract}
Overfitted codecs compress an image by learning a decoder tailored to the
content during the encoding. As such, they trade increased encoding complexity
for strong compression performance and low decoding complexity. This work
introduces Cool-chic 5.0, the latest version in the Cool-chic series of
overfitted codecs, featuring an updated decoder architecture and an improved
optimization process. Cool-chic 5.0 outperforms all overfitted codecs with 10
times less encoding iterations. It offers -11\% rate reduction compared to the
state-of-the-art conventional codec H.266/VVC. It is also competitive with
modern autoencoders such as MLIC++ while featuring a decoding complexity 250
times lower. This work is made open-source at
\url{https://github.com/Orange-OpenSource/Cool-Chic}.
\end{abstract}

\section{Introduction}
\IEEEPARstart{O}{verfitted} codecs such as Cool-chic \cite{coolchic-1-ladune}
are a novel learned compression paradigm. The key idea of overfitted codecs is
to encode an image by training (\textit{i.e.}, overfitting) a decoder and a
latent representation so that they minimize the rate-distortion cost for this
image. The decoder and latent domain are thus tailored to each image, allowing
for compelling compression performance and low decoding complexity.

Similarly to autoencoders \cite{mlicpp-wei,lic-hpcm-li}, overfitted codec are
built with neural networks and learned by optimizing a rate-distortion cost
function. This allows to jointly optimized all the parameters of non-linear
functions, yielding better compression efficiency.

Autoencoders rely exclusively on generalization to compress images. They are
optimized during an \textit{offline} training stage with the objective of being
universal \textit{i.e.,} of offering "good" performance on any possible images.
Then, their parameters are frozen and used as is for the inference stage,
requiring autoencoders to have a substantial amount of parameters in order to
generalize to images not seen during the training stage. Consequently, while
modern autoencoders significantly outperform conventional codecs for still image
coding (MLIC++ \cite{mlicpp-wei}, LIC-HPCM \cite{lic-hpcm-li}), their decoding
complexity is orders of magnitude greater, requiring around 1 million
multiplications to decode a single pixel.
\newline

Overfitted codecs discard generalization in favor of content adaptation. They
carry out an \textit{online} rate-distortion optimization during the encoding,
yielding a latent representation and decoder parameters adapted to each
individual image. Thanks to this adaptation, the decoder complexity can be
greatly reduced. Recent overfitted codecs from the Cool-chic series
\cite{coolchic-3.4-philippe,moric-li,lottery-wu} outperform the state-of-the-art
conventional codec VVC, while requiring around 1\,000 multiplications per
decoded pixel: two to three orders of magnitude lower than their autoencoder
counterparts.

Due to the encoder-side rate-distortion optimization, encoding an image with an
overfitted codec implies an iterative optimization of the decoder and compressed
representation. This is an incentive to design neural architectures and training
processes offering fast and reliable convergence, delivering strong compression
performance with reduced encoding complexity.
\newline

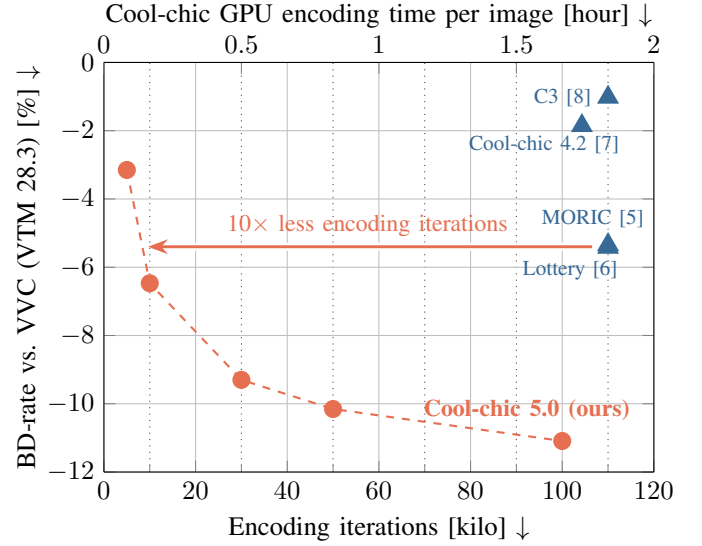
\begin{figure}[t]
    \pgfdeclarelayer{bg}    
    \pgfsetlayers{bg,main}  
    \centering
    \pgfplotsset{minor grid style={dotted,darkgray!50!darkgray}}
    \begin{tikzpicture}
        \begin{axis}[
                grid= major,
                width=\linewidth,
                height=7cm,
                xlabel = {Encoding iterations [kilo] $\downarrow$},
                ylabel = {BD-rate vs. VVC (VTM 28.3) [\%] $\downarrow$} ,
                xmin = 0, xmax = 120, xlabel near ticks, minor x tick num=1,
                ymin = -12, ymax = 0, ylabel near ticks, minor y tick num=0, ytick distance={2},
                enlarge x limits = false, enlarge y limits = false,
                title style={yshift=-0.75ex},
                ylabel shift=-0.15cm,
                grid=both,
                legend style={at={(0.98,0.725)}, anchor= east},
                title style={align=center, font=\small},
                axis x line*=bottom,
            ]

            \addplot[thick, cc_blue, only marks, mark=triangle*, mark size=4pt]
            coordinates {
                (104.3,	-1.852) 
                (110, -1.021) 
                (110, -5.340)  
                (110, -5.418)  
            };

            \addplot[thick, dashed, cc_red, mark=*, mark size=3pt, mark options={solid}] coordinates {
                (5,-3.15)
                (10,-6.47)
                (30,-9.30)
                (50,-10.154)
                (100,-11.091)
                };

            \draw[-{Stealth[length=2.6mm]}, line width=1.1pt, cc_red,
            shorten >=8pt, shorten <=6pt,
            preaction={draw=white, line width=3pt}]
            (axis cs:110,-5.4) -- (axis cs:5,-5.4)
            node[midway, above, font=\small, text=cc_red] {10$\times$ less encoding iterations};

            \node [cc_blue, below, xshift=-0.5cm] at (axis cs:104.3, -1.852){\footnotesize Cool-chic 4.2 \cite{coolchic-open-source-orange}};
            \node [cc_blue, left, xshift=-0.1cm] at (axis cs:110 , -1.021){\footnotesize C3 \cite{c3-kim}};
            \node [cc_blue, above, yshift=0.1cm, xshift=-0.2cm] at (axis cs:110, -5.340){\footnotesize MORIC \cite{moric-li}};
            \node [cc_blue, below, xshift=-0.5cm, yshift=-0.05cm] at (axis cs:110, -5.418){\footnotesize Lottery \cite{lottery-wu}};
            \node [cc_red, above, xshift=-0.5cm, yshift=0.1cm] at (axis cs:100.5, -10.975){\small \textbf{Cool-chic 5.0 (ours)}};
        \end{axis}
        \begin{axis}[
                width=\linewidth,
                height=7cm,
                xmin=0, xmax=2,           
                ymin=0, ymax=100,
                axis x line*=top,
                axis y line=none,
                xlabel={Cool-chic GPU encoding time per image [hour] $\downarrow$},
                x label style={at={(axis description cs:0.5,1.16)}, anchor=south},
            ]
        \end{axis}

    \end{tikzpicture}
    \caption{Compression performance and encoding complexity of overfitted
        codecs on CLIC20 professional validation set \cite{clic20-dataset}.}
    \label{fig:encoder-complexity-bdrate-clic20}
\end{figure}

This paper presents Cool-chic 5.0, the latest iteration in the Cool-chic series.
It proposes improvements on all areas of the codec, refining both the decoder
architecture and the encoding process. In particular, our contributions are as
follows:
\begin{enumerate}[i)]
    \item An entropy model conditioned on multiple latent features;
    \item A linear stabilizer layer speeding-up the convergence;
    \item A refined differentiable proxy for the latent optimization;
    \item A second-order optimizer for the neural networks.
\end{enumerate}
These contributions offer improved compression performance and faster encoding
while maintaining a low decoding complexity. Compared to other overfitted
codecs, Cool-chic 5.0 encodes image 10 times faster for identical compression
performance. Moreover, it achieves 7\% rate reduction at similar
encoding and decoding complexity. This allows Cool-chic 5.0 to obtain 11\% rate
reduction compared to the conventional codec VVC. This is compression
performance comparable to modern autoencoders like MLIC++ despite Cool-chic
decoder being 250 times less complex.

To foster the development of overfitted codecs, Cool-chic 5.0 is made
open-source at \url{https://github.com/Orange-OpenSource/Cool-Chic}.

\section{Related works}


\subsection{Image compression}

Early overfitting-based compression methods emerged from Implicit Neural
Representations (INRs) \cite{inr-sitzmann} which represent data as functions,
modeled using neural networks. COIN \cite{coin-dupont} proposes to adapt INR to
image compression by overfitting a \textit{synthesis} neural network to
represent the image and then compressing its parameters. Subsequent work such as
COIN++ \cite{coinpp-dupont} and Strümpler et al. \cite{inr-strumpler} introduce
meta-learned neural network initialization for faster training.
\newline

COIN-based methods suffers from a lack of local information. Indeed, each
parameter of the synthesis neural network influences all the pixels of the
image, making it hard to offer high image quality for texture-rich contents.
Moreover, efficient entropy modeling of neural network parameters is not
straightforward, leading to important cost of transmission for the neural network
parameters. Cool-chic \cite{coolchic-1-ladune} proposes a solution to both
issues by introducing a set of hierarchical latent grids. These grids are
spatially organized, featuring local information and allowing for easier entropy
modeling with a learned autoregressive probability model.
\newline

The Cool-chic codec has been refined in several follow-up works. Leguay et al.
\cite{coolchic-2-leguay} propose a more expressive synthesis, C3 \cite{c3-kim}
improves the optimization process, Philippe et al. \cite{coolchic-3.4-philippe}
enhance the upsampling of the hierarchical latent grids. More recently, MORIC
\cite{moric-li} proposes spatially adaptive synthesis, LotteryCodec
\cite{lottery-wu} leverages the lottery ticket hypothesis to convey network
parameters more efficiently and Dogaroglu et al.
\cite{multiresolution-context-dogaroglu} enhance the entropy model with
additional context values. Acceleration of the convergence using meta
initialization is explored by Borell et al. \cite{hypercool-pep}.
\newline

Overfitted codecs are also applied to other use-cases. The Cool-chic-based
lossless codec FNLIC \cite{fnlic-zhang} achieves results comparable with modern
lossless codecs such as JPEG-XL. Benjak et al. investigate progressive
\cite{progressive-coolchic-benjak} and scalable \cite{scalable-coolchic-benjak}
image compression using Cool-chic. Lin et al. propose an object-based coding
scheme \cite{object-disentangled-lin} exploiting Cool-chic to compress the
different objects. Ballé et al. focus on subjective visual quality
\cite{ballé2025goodcheapfastoverfitted}, leveraging the Wasserstein distance for
better visual quality, also complementing the decoder with random latent grids.
This method was shown to be successful at the CLIC2025 challenge
\cite{clic2025-coolchic}.

\begin{figure*}[ht]
    \includegraphics[width=\linewidth]{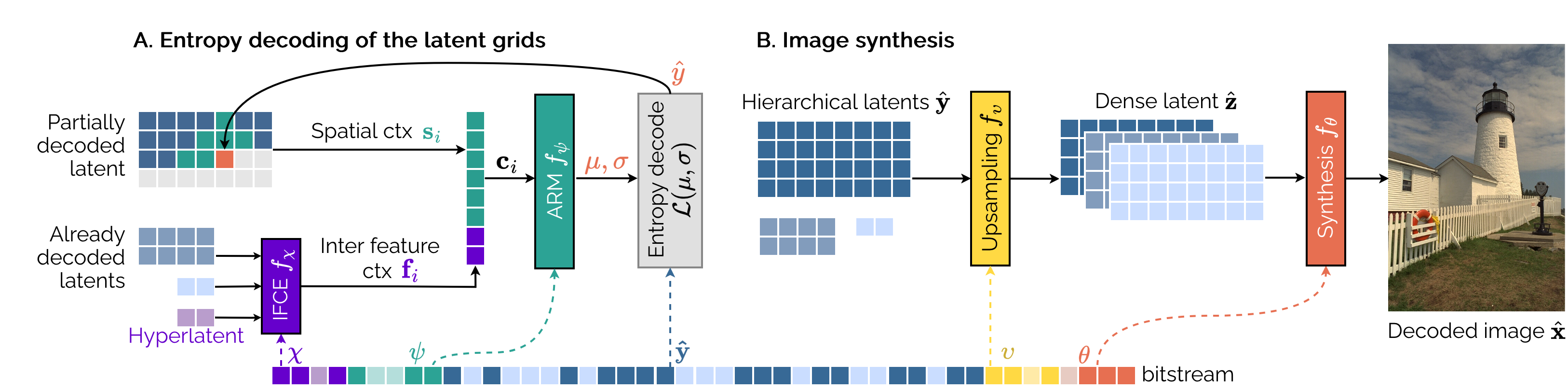}
    \caption{ Overview of the proposed decoder for Cool-chic 5.0, with $L=3$
    latent grids (in blue) and 1 hyperlatent grid (in purple). The hyperlatent
    grids are discarded once step A is completed \textit{i.e.,} they are not
    used to reconstruct the decoded image in step B. IFCE stands for Inter
    Feature Context Extractor. }
    \label{fig:decoder-overview}
\end{figure*}

\subsection{Video compression}

The first overfitted video compression methods extend the ideas of COIN,
representing video as functions modeled using neural networks. The NeRV
\cite{nerv-chen} family of codecs (HNeRV \cite{hnerv-chen}, HiNeRV
\cite{hinerv-kwan}, FFNeRV \cite{ffnerv-lee}, NVRC \cite{nvrc-kwan}) have been
continuously refining this paradigm. These methods represent videos as neural
network parameters and often encode dozens or hundreds of video frames together,
mutualizing parameters across time. This brings important latency, preventing
the usage of low-delay coding configurations. Despite compelling compression
performances, this family of method tends to exhibit substantial decoding
complexity. For instance, NVRC decoding complexity ranges from 100\,000 to 1
million multiplications per pixel depending on the target quality.
\newline

Methods with prominent latent representation such as Cool-chic address these two
issues. They perform sequential encoding of successive frames, enabling
low-delay coding configurations. They also feature low decoding complexity,
typically around 1\,000 multiplications per pixel. C3 \cite{c3-kim} modifies
Cool-chic autoregressive model to take into account spatiotemporal context.
Other approaches rely on Cool-chic to transmit motion information and residue
\cite{coolchic-video-3.1-leguay}. This scheme is extended to exploit pre-trained
optical flow estimators as guidance during the encoding
\cite{coolchic-video-4-leguay}. CNVC \cite{cnvc-li} explores similar ideas and
also introduces an additional entropy constraint on the neural network
parameters as well as learned module to enhance the quality of the temporal
prediction.

\section{The Cool-chic decoder}

This section explains the decoding of an image with Cool-chic 5.0. It is
composed of 2 main steps, illustrated in Fig. \ref{fig:decoder-overview}. First,
the hierarchical latent representation is entropy decoded using a neural network
as probability model (section \ref{sec:entropy-decoding}). Then, the decoded
image is reconstructed from the hierarchical latent representation (section
\ref{sec:reconstructing-the-image}).

\subsection{Entropy decoding of the latent grids}
\label{sec:entropy-decoding}

\subsubsection{Auto-regressive probability model} Cool-chic latent representation $\qlatent$ is a set of
$\nlatentgrids$ discrete, hierarchical, two-dimensional grids defined as:
\begin{equation}
    \qlatent = \{ \qlatent^{(0)}, \ldots, \qlatent^{(\nlatentgrids - 1)}\},\
    \qlatent^{(k)} \text{ shape is } (\tfrac{H}{2^k}, \tfrac{W}{2^k}),
\end{equation}
where $H$ and $W$ denote the height and width of the image.
\newline

The latent representation is transmitted using an entropy coding algorithm
(arithmetic coding), requiring to model its probability distribution. Here, a
factorized model is used, conditioning the distribution of each latent value
$\hat{y}_i$ on a causal context vector $\mathbf{c}_i$ composed of already
decoded values:
\begin{equation}
    p(\qlatent) = \prod_i p(\hat{y}_i \mid \mathbf{c}_i).
    \label{eq:factorized-probability}
\end{equation}

The probability of each scalar latent value $\hat{y}_{i} \in \mathbb{Z}$
is computed by integrating the Probability Density Function (PDF) $f$, following
a Laplace distribution:
\begin{equation}
    p(\hat{y}_{i} \mid \mathbf{c}_i) = \int_{\hat{y}_{i} - 0.5}^{\hat{y}_{i}+0.5} f(x)\mathrm{d}x,
    \text{ with } f \sim \mathcal{L}(\mu_i, \sigma_i).
    \label{eq:laplace-distribution}
\end{equation}

\begin{figure}
    \includegraphics[width=\linewidth]{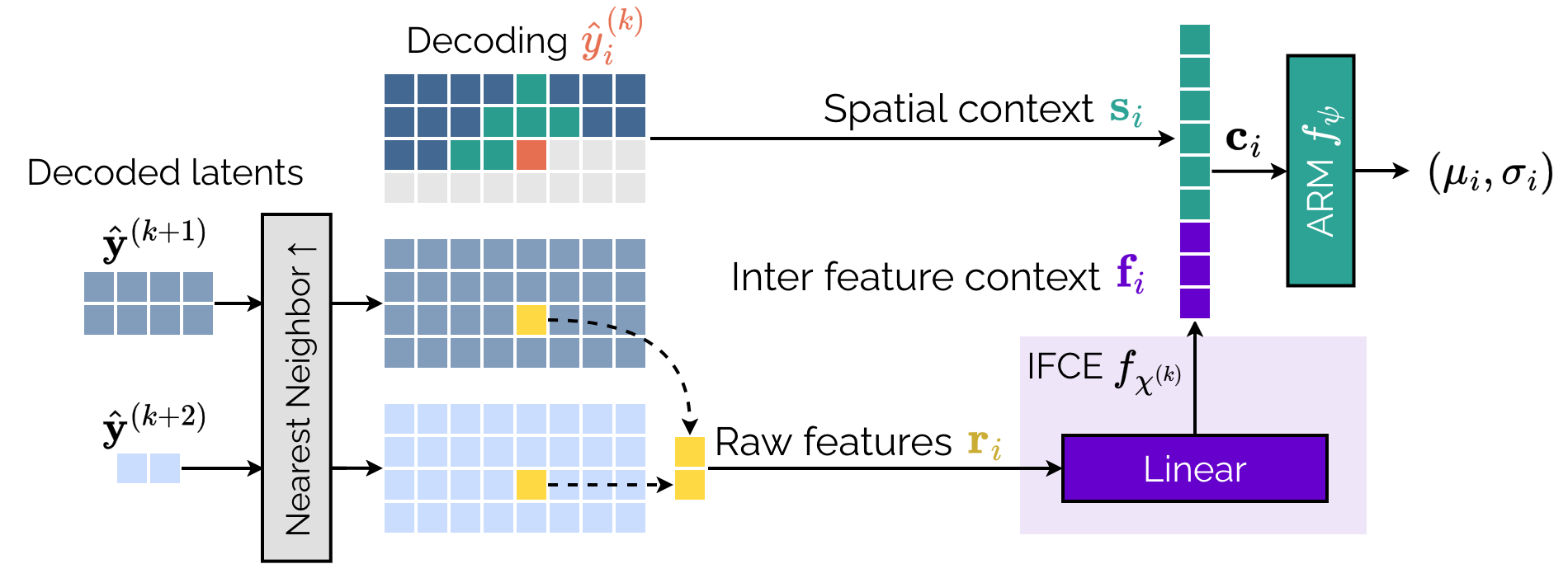}
    \caption{Autoregressive probability model (ARM) and Inter Feature Context
    Extractor (IFCE) to compute the distribution parameters $(\mu_i, \sigma_i)$
    used for the entropy coding of $\hat{y}^{(k)}_i$.
    }
    \label{fig:IFCE}
\end{figure}

Each latent value $\hat{y}_{i}$ has its own probability distribution,
parameterized by its expectation $\mu_i$ and standard deviation $\sigma_i$.
These parameters are obtained by an Auto-regressive Module (ARM) neural network
$\arm$ applied on the context $\mathbf{c}_i$, as shown in Fig. \ref{fig:IFCE}.
The context $\mathbf{c}_i$ is a concatenation of two elements. First, a spatial
context $\mathbf{s}_i$ made of causal neighboring values from the current latent
grid. Then, an inter feature context $\mathbf{f}_i$ carrying information from
already decoded latent grids. The distribution parameters are thus computed as
follows:
\begin{equation}
    \mu_i, \sigma_i = \arm(\mathbf{c}_i), \text{ with } \mathbf{c}_i = \mathtt{concat}(\mathbf{s}_i, \mathbf{f}_i).
\end{equation}

\subsubsection{Inter feature context extractor} The latent grids are decoded
successively starting with the lowest resolution $ \qlatent^{(\nlatentgrids -
1)}$ up to the highest one $\qlatent^{(0)}$, allowing to leverage redundancies
between the latent grids. We propose to enrich the probability model of some of
the latent grids $\qlatent^{(k)}$ using a dedicated \textit{Inter Feature
Context Extractor} (IFCE) network $f_{\ifceparam^{(k)}}$, modeling the
statistical dependencies between the current grid and the $\nlatentgrids -k - 1$
already decoded grids. These already decoded grids are first upscaled to match
the resolution of the current grid. Then the pixel located at position $i$ in
all grids are sampled as a one-dimensional vector $\mathbf{r}_i$ and fed to the
IFCE to obtain the inter feature context $\mathbf{f}_i$:
\begin{equation}
    \mathbf{f}_i = f_{\ifceparam^{(k)}}(\mathbf{r}_i).
\end{equation}
We use a nearest neighbor upsampler for IFCE since it requires no
multiplication, offering easy integer arithmetic implementation, a requirement
for cross-platform operability.
\newline

Only the first few latent grids with the highest resolution use the IFCE,
allowing to keep complexity and number of parameters low. Moreover, each IFCE
$f_{\ifceparam^{(k)}}$ is implemented as a single-layer (\textit{i.e.} linear)
neural network to ensure low complexity. For other latent which does not use
IFCEs, the inter feature context $\mathbf{f}_i$ is set to zero.
\newline

\subsubsection{Hyperlatent grids}
\label{sec:hyperlatent}
Hyperprior is a mechanism where latent grids are conveyed solely to help decode
some other latent grids \textit{i.e.,} without being used by the synthesis to
reconstruct the signal. Here, it is proposed to introduce hyperlatent grids
denoted as $\qlatent_h$, a set of hierarchical latent grids:
\begin{equation}
    \qlatent_h = \{ \qlatent^{(b)}_h, \ldots, \qlatent^{(\nlatentgrids_h - 1)}_h\},\
    \qlatent^{(k)}_h \text{ shape is } (\tfrac{H}{2^{k}}, \tfrac{W}{2^{k}}),
\end{equation}
where $b$ sets the maximal spatial dimension of the hyperlatent grids
\textit{e.g.,} when $b=4$ the biggest hyperlatent grids is of size
$(\tfrac{H}{16}, \tfrac{W}{16})$. The parameter $L_h$ sets the resolution of the
smallest hyperlatent grid. Typically, $\nlatentgrids_h = \nlatentgrids$
\textit{i.e.,} both the hyperlatent and the latent grids have the same lowest
resolution.
\newline

The hyperlatent grids are entropy coded using the same ARM as the other latent
grids. They are leveraged by other latent grids, through their IFCEs, to obtain
a more accurate probability model. The hyperlatent grids $\qlatent_h$ are only
used during the entropy decoding step and discarded before reconstructing the
image from the main latent grids $\qlatent$, limiting the overall decoding
complexity. Figure \ref{fig:decoder-overview} presents the decoding of an image
with three latent grids and one hyperlatent.

\subsection{Image synthesis}
\label{sec:reconstructing-the-image}

\begin{figure}[t]
    \centering
    \includegraphics[width=\linewidth]{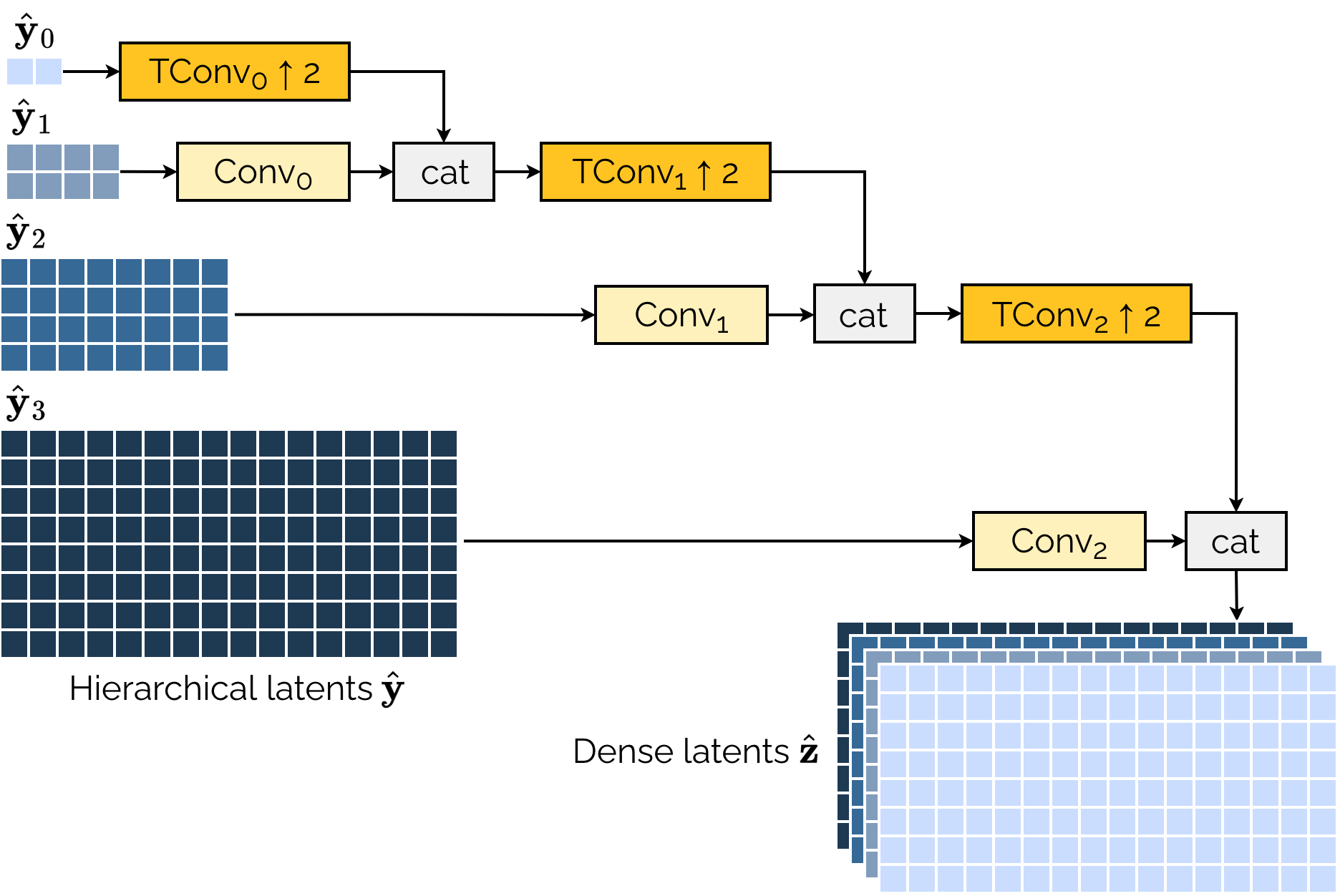}
    \caption{Architecture of the upsampling $\upsample$. Conv and TConv denote
    convolution and transposed convolution layers.}
    \label{fig:upsampling}
\end{figure}

Once the latent grids are entropy decoded, they are used to reconstruct the
image. Two neural networks compute the decoded image $\codedimg$. First, a
neural upsampling $\upsample$ maps the hierarchical latent grids $\qlatent$ to a
dense representation $\denselatent$:
\begin{equation}
    \denselatent = \upsample(\qlatent), \text{ where } \denselatent \text{ shape is } (\nlatentgrids, H, W).
\end{equation}
The upsampling network is identical to the one proposed by Philippe et al.
\cite{coolchic-3.4-philippe}, shown in Fig. \ref{fig:upsampling}. It leverages
several symmetrical and separable convolution kernels to maintain a low decoding
complexity.
\newline

\begin{figure*}[t]
    \centering
    \begin{subfigure}{0.495\linewidth}
        \includegraphics[width=\linewidth]{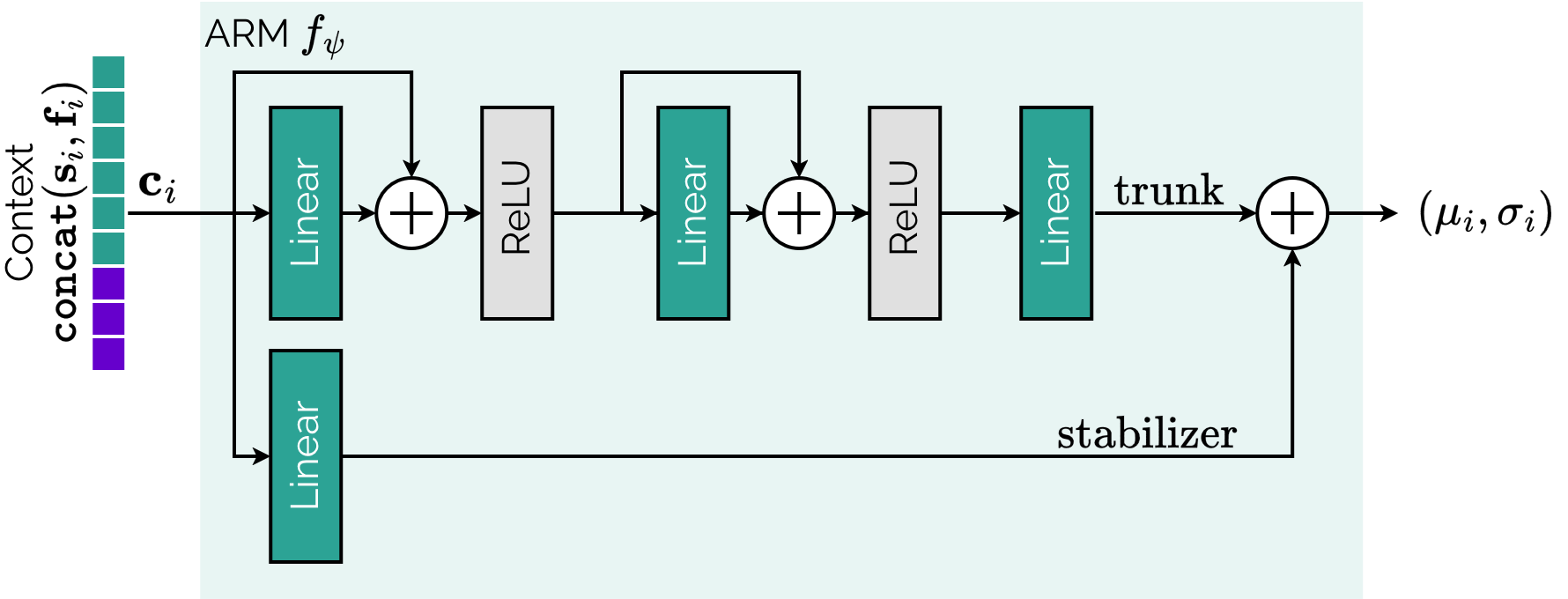}
        \caption{Auto regressive probability model (ARM) $\arm$.}
        \label{fig:arm-detailed}
    \end{subfigure}
    \begin{subfigure}{0.495\linewidth}
        \includegraphics[width=\linewidth]{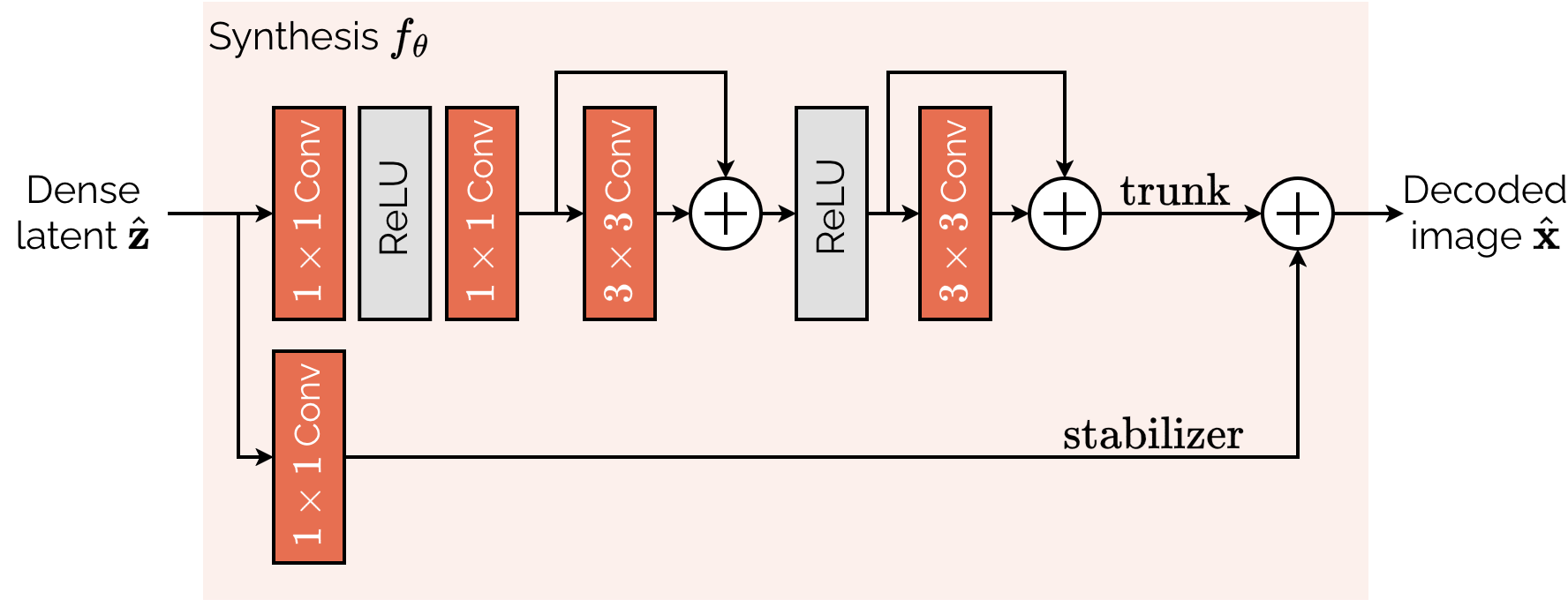}
        \caption{Synthesis transform $\synth$.}
        \label{fig:synthesis}
    \end{subfigure}
    \caption{Architecture of the ARM $\arm$ and Synthesis $\synth$
    neural networks. Both networks feature the linear residual stabilizer.}
    \label{fig:stabilizer}
\end{figure*}

Then, the synthesis transform $\synth$ maps the dense representation to the
decoded image:
\begin{equation}
    \codedimg = \synth(\denselatent), \text{ where } \codedimg \text{ shape is } (C, H, W),
\end{equation}
where $C$ is the number of image channels, typically 3 for RGB. The synthesis
architecture is presented in Fig. \ref{fig:synthesis}. The trunk branch is
composed of point-wise ($1\times 1$ kernel) convolution layers mapping the $L$
input features to $C$ channels. Then, successive residual layers act as a
post-processing step to enhance the quality of the image.


\subsection{Linear residual stabilizer}
\label{sec:linear-stabilizer}
To improve the convergence, it is proposed to complement the synthesis and the
ARM neural networks with a \textit{stabilizer} branch. The stabilizer is a
residual linear layer, operating in parallel to the \textit{trunk} branch.
Figure \ref{fig:stabilizer} shows the linear residual stabilizer applied to the
synthesis $\synth$ and the ARM $\arm$. Denoting $f$ the function computed by a
neural network, it is decomposed as follows:
\begin{equation}
    f(\mathbf{x}) = \mathrm{trunk}(\mathbf{x}) + \mathrm{stabilizer}(\mathbf{x}).
\end{equation}

\section{Encoding an image with Cool-chic}

This section presents the encoding process of Cool-chic which is build on a
gradient-based minimization of each image rate-distortion cost, requiring
end-to-end differentiable operations. This optimization process features
successive stages and ends by the coding of the neural network parameters.

\subsection{Encoding-time optimization of the rate-distortion cost}

Cool-chic encodes an image by learning the decoder neural networks
$(f_{\armparam}, f_{\ifceparam}, f_{\upparam}, f_{\synthparam})$ and the latent
grids $\qlatent$ offering the best rate-distortion trade-off. For the sake of
clarity, since both the main latent representation $\qlatent$ and the hyper
latent representation $\qlatent_h$ are optimized identically, we denote the set
of all latent grids as $\qlatent$. The goal of the encoding is thus to minimize
the following rate-distortion cost:
\begin{equation}
    \qlatent, \armparam, \ifceparam, \upparam, \synthparam
    = \arg\min \mathrm{D}(\img, \codedimg) + \lambda \mathrm{R}(\qlatent).
    \label{eq:training-loss}
\end{equation}
Here, $\mathrm{D}$ measures the mean squared error between the original and
compressed images. $\mathrm{R}$ corresponds to the rate of the latent
grids, estimated through its cross entropy:
\begin{equation}
    \mathrm{R}(\qlatent) = -\log_2 p(\qlatent),
\end{equation}
with the latent distribution modeled following eq.
\eqref{eq:factorized-probability} and \eqref{eq:laplace-distribution}.

\subsection{Gradient-based optimization with a second-order method}

The rate-distortion objective function presented in eq. \eqref{eq:training-loss}
is minimized iteratively using a gradient-based method. The parameters to
optimize through gradient descent are of two different natures. On one hand,
there are the neural network parameters $(f_{\armparam}, f_{\ifceparam},
f_{\upparam}, f_{\synthparam})$ which typically represents around 2\,000
parameters. On the other hand, there are the latent grids which are composed of
1\,000\,000 parameters for high-resolution images.
\newline

Cool-chic 5.0 uses the second-order optimizer SOAP \cite{vyas-soap} for the
neural network parameters, while resorting to the first-order Adam optimizer
\cite{kingma-adam} for the latent grids. Crucially, this allows to benefit of
the better convergence of the SOAP optimizer with a minimal computational
overhead since it is applied on few parameters.

\subsection{Improved differentiable proxy for quantization}

Optimizing the training objective presented in eq. \eqref{eq:training-loss} with
a gradient-based method requires all optimized quantities to be real-valued. As
such, a \textit{continuous} latent representation $\latent$ is optimized and
quantized to the \textit{discrete} representation $\qlatent$:
\begin{equation}
    \qlatent = Q(\latent), \text{ with } Q \text{ a uniform scalar quantizer.}
\end{equation}

However, the gradient of the quantization operation with respect to its input is
zero nearly everywhere, cancelling all gradients for upstream variables. In order to
allow gradient back propagation to the continuous latent representation
$\latent$, quantization is replaced by a differentiable proxy $Q_{train}$ during
training \cite{c3-kim}:
\begin{equation}
    \qlatent_{train} = Q_{train}(\latent) = s_T(s_T(\latent) + \mathbf{n}_{\sigma}).
    \label{eq:q-train}
\end{equation}

This differentiable proxy is composed of two elements. First, a softround
function $s_T$, a softer version of the (hard) rounding operation, parameterized
by a temperature parameter $T$. Then an independent zero-mean noise vector
$\mathbf{n}_{\sigma}$ with the same shape as the latent representation, parameterized by
its standard deviation $\sigma$. The softround function $s_T$ is defined
following Agustsson and Theis \cite{universal-quantization-agustsson}:
\begin{equation}
    s_T(x) = \lfloor x \rfloor + \frac{\tanh(\Delta / T)}{2\tanh(1/{2T})} + \frac{1}{2}
    \text{ and } \Delta = x - \lfloor x \rfloor - \frac{1}{2}.
    \label{eq:softround}
\end{equation}
The temperature parameter $T$ is decreased during training so that the softround
function is similar to the actual hardround function at the end of the training.

The additive noise vector $\mathbf{n}_{\sigma}$ follows a centered Gaussian distribution.
Its standard deviation is also decreased during training so that the
quantization proxy (eq. \eqref{eq:q-train}) resembles the hard quantization
towards the final steps of the training.

\subsection{Multi-stage encoding}

Cool-chic encoding implements a 3-stage training process, composed of a warm-up,
a main stage and a hard round stage. All stages optimize the training loss
presented in eq. \eqref{eq:training-loss}.
\newline

\subsubsection{Warm-up} The warm-up selects the most suited initialization among
a set of randomly initialized candidates. To do so, $N$ candidates are
initialized and trained during a few hundred iterations. Then, the 2 candidates
achieving the best rate-distortion cost are further refined. Finally, the best
one is selected for the main training stage.
\newline

\subsubsection{Main stage} This stage represents most of the training process.
Several training hyperparameters are continuously scheduled during this stage.
The random noise standard deviation (eq. \eqref{eq:q-train}) is decreased, as
well as the temperature of the softround function (eq. \eqref{eq:softround}). As
such, the proxy quantization function used during training gradually change
towards the rounding function. The learning rate is also decreased following a
cosine scheduling following \cite{c3-kim}.
\newline

\subsubsection{Hardround stage} The last few hundred iterations use the actual
quantization during the forward pass so that the decoder is prepared to operate
on integer values. During this step, only the neural network parameters are
optimized \textit{i.e.,} the latent grids remain unchanged.

\subsection{Neural network transmission}
\label{sec:nn-signaling}

The last step of the encoding is the compression of the neural network
parameters. Neural networks parameters have a different nature than latent
variables, featuring smaller amplitudes and no spatial organization. There are also
significantly less neural networks parameters (typically 2\,000) than latent
values (1\,000\,000 for a high-resolution image). Consequently, the neural
networks are conveyed with a separate quantization and signaling scheme.
\newline

\subsubsection{Coding scheme}
Weights and biases of the neural networks composing the decoder are real-valued
during the training process. To send them efficiently, an entropy coding
algorithm is used, requiring the parameters to be discrete. A pair of
quantization steps $\Delta \in \mathbb{R}^2$ (weights and biases) is used for
each of the 4 submodules composing the decoder, namely the IFCE $f_\chi$, the
ARM $\arm$, the upsampling $\upsample$ and the synthesis $\synth$. That is the
quantization steps of the neural networks are:
\begin{equation}
    \Delta_{NN} = \{
        \Delta_{\armparam},
        \Delta_{\synthparam},
        \Delta_{\upparam},
        \Delta_{\chi}
    \}.
\end{equation}

Once the neural network parameters are quantized to discrete values, they are
coded using an Exp-Golomb code as the parameters tend to follow a Laplace
distribution. This also allows transmitting parameters without restricting their
dynamic. Similarly to quantization steps, a pair of Exp-Golomb orders $\kappa
\in \mathbb{N}^2$ are used (weights and biases) for each of the 4 submodules
composing the decoders:
\begin{equation}
    \kappa_{NN} = \{
        \kappa_{\armparam},
        \kappa_{\synthparam},
        \kappa_{\upparam},
        \kappa_{\chi}.
    \}
\end{equation}

\subsubsection{Parameters selection}

The selection of the quantization steps $\Delta_{NN}$ and Exp-Golomb orders
$\kappa_{NN}$ is done successively for each of the 4 submodules by evaluating a
list of pre-defined candidates. The list of possible values for each parameter
is kept small to limit the computational overhead \textit{e.g.,} 8 possible
quantization steps ranging from $2^{-8}$ to $2^{-1}$.
\newline

The optimal quantization and Exp-Golomb parameters are selected based on the
rate-distortion cost they offer:
\begin{equation}
    \Delta_{NN}, \kappa_{NN}
    = \arg\min \mathrm{D}(\img, \codedimg) + \lambda (\mathrm{R}(\qlatent) + \mathrm{R}_{NN}),
\end{equation}
with $\mathrm{R}_{NN}$ the rate of the neural network. Note that for each
quantization step, it is necessary to decode the image with the quantized
decoder to evaluate the degradation due to the loss of accuracy in the neural
network parameters. Finally, The selected quantization and Exp-Golomb parameters
are transmitted alongside the quantized weights and biases.

\section{Contributions and prior work}

This section highlights our contributions to the decoder architecture and to the
encoder training process. In particular, we discuss the novelty of the proposed
design compared to several prior work.

\subsection{Decoder architecture}

The decoding pipeline of this paper follows the one introduced in the original
Cool-chic paper \cite{coolchic-1-ladune} and later refined in several follow-up
works \cite{coolchic-2-leguay,c3-kim,coolchic-3.4-philippe}. This work
introduces three key improvements to the decoder architecture: the inter feature
context extractor (IFCE) module, the hyperlatent grids and the stabilizer layer.
\newline

\subsubsection{Inter feature context extractor (IFCE)} The idea of leveraging
inter-feature redundancies is already found in autoencoder-based codecs
\cite{autoregressive-minnen} and has thus been adapted to overfitted approaches.
C3 \cite{c3-kim} proposes to condition the decoding of the $k$-th latent grid on
a single $(k+1)$-th latent, limiting the expressivity of the model. In this
work, we propose to leverage all the information available \textit{i.e.}, all
already decoded latent grids. Dugaroglu et al.
\cite{multiresolution-context-dogaroglu} propose to exploit information from all
previously decoded latent grids. However, this is achieved without the IFCE
networks \textit{i.e.} entering directly the latent values into the ARM module,
resulting in less adaptability for the model.
\newline

\subsubsection{Hyperlatent grids} One significant milestone in the development
of autoencoders is the introduction of the hyperprior mechanism
\cite{variational-hyperprior-balle} \textit{i.e.,} auxiliary information whose
sole role is to help in the probability modeling of the other latents.
Crucially, this hyperprior information is not used to synthesize the decoded
image and is discarded once all latent variables are entropy decoded. This work
is the first to extend this mechanism to overfitted codecs allowing for better
compression performance.
\newline

\subsubsection{Stabilizer layer} The proposed stabilizer consists in a single
linear layer operating in parallel to several non-linear layers. It resembles
the residual architecture introduced by ResNet \cite{resnet-he} which improves
the convergence of neural networks. Here, we propose that the residual branch
also performs a linear transform, projecting the latent domain back to the image
domain (for the synthesis) or mapping the context to the Laplace distribution
parameters (for the ARM).

\subsection{Encoding process}

Several works \cite{coolchic-2-leguay,c3-kim,ballé2025goodcheapfastoverfitted}
hint that refining the encoding stage (\textit{i.e.,} quantization proxy
function, hyperparameters, optimizers \textit{etc.}) is crucial for overfitted
codecs and offers significant performance improvement. One contribution of this
work is to conduct extensive experiments to properly study the impact of most of
the training parameters.
\newline

\subsubsection{Quantization proxy function} Differentiable proxies for the
quantization function have been widely studied in the learned compression
literature. Some early works rely on the straight-through approximation
\cite{lossy-autoencoder-theis} while others \cite{variational-hyperprior-balle}
propose to simulate quantization through the addition of a uniform noise.
Soft-then-hard approaches
\cite{soft-then-hard-guo,universal-quantization-agustsson} are also introduced
to better replicate the quantization process. In particular, Agustsson et al.
\cite{universal-quantization-agustsson} motivates the usage of both uniform
random noise and soft rounding. This is applied by C3 \cite{c3-kim} which
samples the noise from a Kumaraswamy distribution.

This work follows the same approach as C3 but carefully reconsider all
hyperparameters related to the relaxed quantization. The softround temperature
$T$ is scheduled so that it is smoother at the beginning of the training and
more closely resembles the actual quantization towards the end of the training.
The random noise is now sampled from a Gaussian distribution. Crucially, while
C3 increases the noise energy during the training, it is instead proposed to
decrease its energy as the training progresses. The rationale is that
high-energy noise fosters the exploration of the parameters space during the
first iterations, but it should then be reduced for the quantization proxy to
mimic the hardround function. These changes lead to better and faster
convergence.
\newline

\begin{table}[t!]
    \caption{Proposed encoder parameters for Cool-chic 5.0.}
    \centering
    \small
    \begin{tblr}{
        colspec={Q[l] Q[l]},
        cell{1}{1}={c=2,r=1}{c,font=\itshape, darkgray!10},
        cell{7}{1}={c=2,r=1}{c,font=\itshape, darkgray!10},
        cell{12}{1}={c=2,r=1}{c,font=\itshape, darkgray!10},
    }
    Number of iterations \\
    \cmidrule{1-Z}
    Total                                               & $100\, 000$                                       \\
    Warm-up candidates                                  & $5$ (\textit{round 1}) then $2$ (\textit{round 2})\\
    Warm-up itr per candidate                           & $400$                                             \\
    Main stage iterations                               & $96\, 700$                                        \\
    Hardround iterations                                & $500$                                             \\
    Dynamic parameters during main stage \\
    \cmidrule{1-Z}
    Learning rate                                       & $10^{-2} \rightarrow 10^{-5}$ (cosine scheduling) \\
    Noise standard deviation                            & $0.22 \rightarrow 0.15$ (linear scheduling)       \\
    Softround temperature $T$                           & $0.35 \rightarrow 0.08$ (linear scheduling)       \\
    Noise type $\mathbf{n}_{\sigma}$                    & Gaussian                                          \\
    Rate constraints \\
    \cmidrule{1-Z}
    Rate-distortion constraints $\lambda$               & $0.02, 0.004, 0.001, 0.0004, 0.0001$              \\
    \end{tblr}
    \label{table:encoder-parameters}
\end{table}

\begin{table}[t!]
    \caption{Proposed decoder parameters for Cool-chic 5.0. The number and
    dimension of the latent grids are conditioned on the number of pixels $HW$ in
    the image. (\textbf{S}) indicates parameters for small images \textit{i.e.,}
    with less than $HW = 10^6$ pixels, while (\textbf{B}) corresponds to bigger
    images.}
    \centering
    \small
    \begin{tblr}{
        colspec={Q[l] Q[c, wd=8mm] Q[c, wd=8mm] Q[c, wd=8mm] Q[c, wd=8mm]},
        cell{1}{1}={c=5,r=1}{c,font=\itshape, darkgray!10},
        cell{5}{1}={c=5,r=1}{c,font=\itshape, darkgray!10},
        cell{6}{2}={c=4,r=1}{l},
        cell{7}{2}={c=4,r=1}{l},
        cell{8}{1}={c=5,r=1}{c,font=\itshape, darkgray!10},
        cell{9}{2}={c=4,r=1}{l},
        cell{14}{1}={c=5,r=1}{c,font=\itshape, darkgray!10},
        cell{15}{2}={c=4,r=1}{c},
    }
    Decoder configurations \\
    \cmidrule{1-Z}
    Configuration name                      & LOP  & MOP & HOP & VHOP \\
    Complexity [kMAC/pix]                   & 0.5  & 1.0 & 2.0 & 3.0  \\
    Parameters [kilo]                       & 0.5  & 1.0 & 1.9 & 2.7  \\
    Latent grids \\
    \cmidrule{1-Z}
    Latent resolutions & (\textit{All configs}) $\tfrac{1}{1}$ to $\tfrac{1}{64}$ (\textbf{S}) or $\tfrac{1}{128}$ (\textbf{B})\\
    Hyperlatent resolutions & (\textit{All configs}) $\tfrac{1}{16}$ to $\tfrac{1}{64}$ (\textbf{S}) or $\tfrac{1}{128}$ (\textbf{B})\\
    Entropy model \\
    \cmidrule{1-Z}
    IFCEs                                   & (\textit{All configs}) For the $\tfrac{1}{1}$, $\tfrac{1}{2}$, $\tfrac{1}{4}$ grids \\
    Inter-feature context $\mathbf{f}$      & 2   & 4   & 6   & 6   \\
    Spatial context $\mathbf{s}$            & 6   & 10  & 14  & 20  \\
    Hidden layers ARM                       & 2   & 2   & 2   & 2   \\
    Layer width ARM                         & 8   & 10  & 20  & 26  \\
    Synthesis \\
    \cmidrule{1-Z}
    Input latent features                   & (\textit{All configs}) 7 (\textbf{S}) or 8 (\textbf{B}) \\
    Features $1\times 1$ conv.              & 8, 3 & 16, 3  & 48, 3  & 64, 3    \\
    $3\times 3$ Conv. post filters          & 1    & 2      & 2      & 2        \\
    \end{tblr}
    \label{table:decoder-parameters}
\end{table}

\subsubsection{Gradient-based optimization} Overfitted codecs rely on
first-order gradient-based estimator that is the Adam optimizer
\cite{kingma-adam} to obtain parameters minimizing the rate-distortion cost. In
the context of autoencoders, second-order optimizer involving an estimate of
loss function curvature are shown to offer better and faster convergence
\cite{leveraging-second-order-zhang}. In this work, the second-order optimizer
SOAP \cite{vyas-soap} is used to optimize the neural network parameters,
yielding faster and better convergence with little to no computational overhead
since neural network represents less than 1\% of the number of parameters
optimized. Indeed, the latent grids are composed of much more parameters and are
still learned with the Adam optimizer.

\section{Results}

\newcommand{\yminbd}{-20}
\newcommand{\ymaxbd}{15}
 \newcommand*\redround{\tikz{\node[shape=circle,inner sep=2pt,fill=cc_red]{}}}
 \begin{figure*}[tb]
    \begin{subfigure}[t]{0.5\linewidth}
    \pgfdeclarelayer{bg}    
    \pgfsetlayers{bg,main}  
    \centering
    \begin{tikzpicture}
        \begin{semilogxaxis}[
                grid= major,
                width=\linewidth,
                height=7cm,
                xlabel = {Decoder complexity [MAC / pixel] $\downarrow$},
                ylabel = {BD-rate vs. VVC (VTM 28.3) [\%] $\downarrow$} ,
                xmin = 100, xmax = 1000000, xlabel near ticks, minor x tick num=0,
                ymin = \yminbd, ymax = \ymaxbd, ylabel near ticks, minor y tick num=0, ytick distance={5},
                enlarge x limits = true, enlarge y limits = false,
                title style={yshift=-0.75ex},
                ylabel shift=-0.15cm,
                legend columns=2,
                legend style={at={(0.5,1.02)}, anchor=south,
                    /tikz/column 2/.style={
                        column sep=15pt,
                    },
                },
                legend cell align={left},
            ]


            \addplot[thick, cc_blue, only marks, mark=triangle*, mark size=4pt, ]
            coordinates {
                (1433,	-1.852) 
                (2925 , -1.021) 
                (2891, -5.340)  
                (3653, -5.418)  
            };
            \addlegendentry{\small \redround  \ Overfitted codecs}

            \addplot[thick, dashed, cc_red, mark=*, mark size=3pt, mark options={solid}, forget plot] coordinates {
                (493, 0.692)
                (1056, -7.338)
                (1991, -11.091)
                (2936, -11.85)
            };

            \addplot[thick, cc_green, only marks, mark=square*, mark size=3pt] coordinates {
                (80000,	46.565) 
                (350000, 12.43) 
                (390000, -1.963) 
                (816000, -12.025) 
                (362000, -8.57) 
                (1007000, -16.983) 
            };
            \addlegendentry{\small Autoencoders}

            \node [cc_green, left, xshift=-0.1cm] at (axis cs:350000, 12.43){\footnotesize Cheng \cite{autoencoder-2020-cheng}};
            \node [cc_green, above, yshift=0.05cm] at (axis cs:390000, -1.963){\footnotesize ELIC \cite{elic-2022-he}};
            \node [cc_green, left, xshift=-0.1cm] at (axis cs:816000, -12.025){\footnotesize MLIC++ \cite{mlicpp-wei}};
            \node [cc_green, above, yshift=0.05cm] at (axis cs:362000, -8.57){\footnotesize DCVC-RT \cite{dcvc-rt-jia}};
            \node [cc_green, above, yshift=0.05cm, xshift=-0.4cm] at (axis cs:1007000, -16.983){\footnotesize LIC HPCM \cite{lic-hpcm-li}};

            \node [cc_blue, right, xshift=0.05cm, yshift=-0.15cm] at (axis cs:1433,	-1.852){\footnotesize Cool-chic 4.2 \cite{coolchic-open-source-orange}};
            \node [cc_blue, right, xshift=0.1cm] at (axis cs:2925 , -1.021){\footnotesize C3 \cite{c3-kim}};
            \node [cc_blue, right, xshift=0.1cm, yshift=0.10cm] at (axis cs:2891, -5.340){\footnotesize MORIC \cite{moric-li}};
            \node [cc_blue, right, xshift=0.1cm, yshift=-0.15cm] at (axis cs:3653, -5.418){\footnotesize Lottery \cite{lottery-wu}};

            \node [cc_red, below, yshift=-0.2cm, xshift=0.1cm] at (axis cs:1986, -11.091){\small \textbf{Cool-chic 5.0 (ours)}};

        \end{semilogxaxis}
    \end{tikzpicture}
    \caption{CLIC20 pro validation dataset.}
    \label{fig:decoder-complexity-bdrate-clic20}
    \end{subfigure}
    \begin{subfigure}[t]{0.5\linewidth}
    \centering
    \begin{tikzpicture}
        \begin{semilogxaxis}[
                grid= major,
                width=\linewidth,
                height=7cm,
                xlabel = {Decoder complexity [MAC / pixel] $\downarrow$},
                ylabel = {BD-rate vs. VVC (VTM 28.3) [\%] $\downarrow$} ,
                xmin = 100, xmax = 1000000, xlabel near ticks, minor x tick num=0,
                ymin = \yminbd, ymax = \ymaxbd, ylabel near ticks, minor y tick num=0, ytick distance={5},
                enlarge x limits = true, enlarge y limits = false,
                title style={yshift=-0.75ex},
                ylabel shift=-0.15cm,
                legend columns=2,
                legend style={at={(0.5,1.02)}, anchor=south,
                    /tikz/column 2/.style={
                        column sep=15pt,
                    },
                },
                legend cell align={left},
            ]

            \addplot[thick, dashed, cc_red, mark=*, mark size=3pt, mark options={solid}, forget plot] coordinates {
                (476, 5.892)
                (1030, -0.418)
                (1937, -2.640)
            };

            \addplot[thick, cc_blue, only marks, mark=triangle*, mark size=4pt]
            coordinates {
                (1433, 2.839) 
                (2626, 8.017) 
                (2665, 0.322)  
                (3523, 5.149)  
            };
            \addlegendentry{\small\redround \ Overfitted codecs}

            \addplot[thick, cc_green, only marks, mark=square*, mark size=3pt] coordinates {
                (80000,	36.127) 
                (350000, 9.297) 
                (390000, -2.175) 
                (816000, -12.100) 
                (362000, -10.427) 
                (1007000, -17.428) 
            };
            \addlegendentry{\small Autoencoders}

            \node [cc_green, above, yshift=0.05cm] at (axis cs:350000, 9.297){\footnotesize Cheng \cite{autoencoder-2020-cheng}};
            \node [cc_green, above, yshift=0.05cm] at (axis cs:390000, -2.175){\footnotesize ELIC \cite{elic-2022-he}};
            \node [cc_green, left, xshift=-0.1cm, yshift=-0.05cm] at (axis cs:816000, -12.100){\footnotesize MLIC++ \cite{mlicpp-wei}};
            \node [cc_green, left, xshift=-0.1cm] at (axis cs:362000, -10.427){\footnotesize DCVC-RT \cite{dcvc-rt-jia}};
            \node [cc_green, left, xshift=-0.1cm] at (axis cs:1007000, -17.428){\footnotesize LIC HPCM \cite{lic-hpcm-li}};

            \node [cc_blue, right, xshift=0.1cm] at (axis cs:1433, 2.839){\footnotesize Cool-chic 4.2 \cite{coolchic-open-source-orange}};
            \node [cc_blue, right, xshift=0.1cm] at (axis cs:2626, 8.017){\footnotesize C3 \cite{c3-kim}};
            \node [cc_blue, right, xshift=0.1cm] at (axis cs:2665, 0.322){\footnotesize MORIC \cite{moric-li}};
            \node [cc_blue, right, xshift=0.1cm] at (axis cs:3523, 5.149){\footnotesize Dogaroglu \cite{multiresolution-context-dogaroglu}};

            \node [cc_red, below, yshift=-0.2cm, xshift=0.1cm] at (axis cs:1924, -2.64){\small \textbf{Cool-chic 5.0 (ours)}};

        \end{semilogxaxis}
    \end{tikzpicture}
    \caption{Kodak dataset.}
    \label{fig:decoder-complexity-bdrate-kodak}
    \end{subfigure}
    \caption{BD-rate versus VVC against the decoding complexity in
    multiplication accumulations (MAC) per pixel.}
    \label{fig:decoder-complexity-bdrate}
\end{figure*}
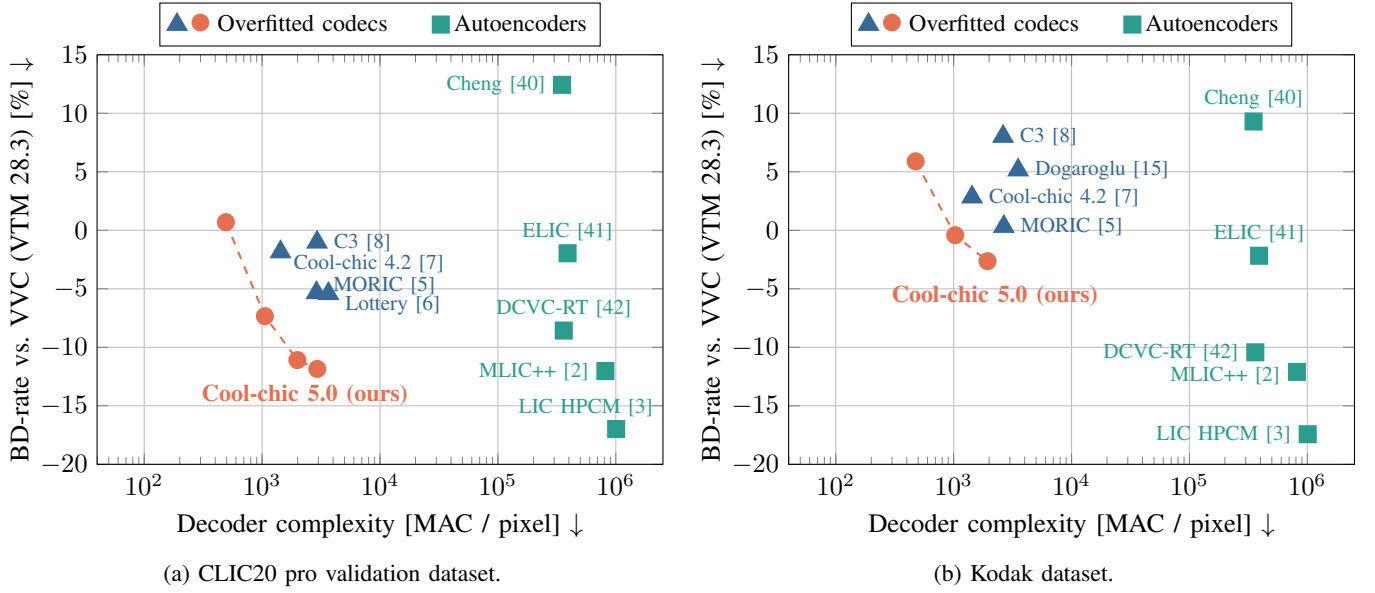

\subsection{Rate-distortion performance}

The proposed Cool-chic 5.0 is evaluated on the Kodak \cite{kodak-dataset} and
CLIC20 pro validation \cite{clic20-dataset} datasets. Kodak contains 24 natural
images of size $512 \times 768$ and CLIC20 contains 41 natural images whose
resolution ranges from $384 \times 512$ to $1370\times 2048$. The quality of the
compressed image is measured with PSNR in the RGB domain. Cool-chic
performance is compared with several anchors: conventional codec (H.266/VVC
\cite{overview-vvc-bross} through VTM 28.3), autoencoders (Cheng et al.
\cite{autoencoder-2020-cheng}, ELIC \cite{elic-2022-he}, DCVC-RT
\cite{dcvc-rt-jia}, MLIC++ \cite{mlicpp-wei}) and other overfitted codecs
(Cool-chic 4.2 \cite{coolchic-open-source-orange}, C3 \cite{c3-kim}, MORIC
\cite{moric-li}, LotteryCodec \cite{lottery-wu} and Dogaroglu et al.
\cite{multiresolution-context-dogaroglu}).
\newline

Table \ref{table:encoder-parameters} presents the encoder parameters used for
Cool-chic 5.0. Several decoder configurations are proposed in Table
\ref{table:decoder-parameters}, with a complexity ranging from 500
multiplications-accumulations (MAC) per decoded pixel to 3\,000 MAC / pixel. The
compression performance of all codecs are represented as a function of their
decoding complexity in Fig. \ref{fig:decoder-complexity-bdrate}.

Cool-chic 5.0 offers the best compression performance from 500 to 3\,000 MAC /
decoded pixel. At equal decoding complexity, it outperforms other overfitted
codecs by up to 7\% \textit{e.g.,} compared to LotteryCodec and MORIC on the
CLIC dataset. Cool-chic 5.0 also achieves 11\% rate reduction compared to VVC.
It is competitive with autoencoders whose decoder is 250 times more complex such
as MLIC++.
\newline

While still outperforming VVC, Cool-chic 5.0 appears to have modest performance
on the Kodak dataset (Fig. \ref{fig:decoder-complexity-bdrate-kodak}), as it is
also the case with other overfitted codecs. This is mostly due to the smaller
compressed file size on this dataset caused by smaller images. Consequently, the
overhead caused by the transmission of the neural network parameters is more
important, hurting the compression performance. This is detailed in section
\ref{sec:rate-nn-experiments}.

\subsection{Encoding complexity}

The main disadvantage of overfitted codecs is their costly encoding stage,
involving the iterative minimization of the image rate-distortion cost shown in
eq. \eqref{eq:training-loss}. Cool-chic 5.0 introduces a number of contributions
to improve the convergence speed:~the addition of linear stabilizer layers,
using a second-order optimizer and a refined differentiable proxy to the
quantization.
\newline

It is proposed to study the compression performance of several overfitted codecs
as a function of the number of encoding iterations. This is obtained by using
the HOP decoder configuration in Table \ref{table:decoder-parameters} and the
encoder parameters presented in Table \ref{table:encoder-parameters}, simply
varying the amount of warm-up and main stage iterations to obtain different
encoding speeds ranging from 5\,000 iterations for the quickest to 100\,000 for
the slowest. Figure \ref{fig:encoder-complexity-bdrate-clic20} shows that
Cool-chic 5.0 outperforms all other overfitted codecs, even with 10 times less
encoding iterations. Indeed, as few as 10\,000 training iterations are required to achieve
a BD-rate of -6.5\% against VVC, while LotteryCodec and MORIC both require
110\,000 iterations to obtain similar performance.

For a CLIC image, performing 1\,000 training iterations requires around 1 minute
on a RTX4090 GPU. As such, Fig. \ref{fig:encoder-complexity-bdrate-clic20} shows
that after 5 minutes of encoding, Cool-chic 5.0 already outperforms VVC.

\subsection{Specific examples}

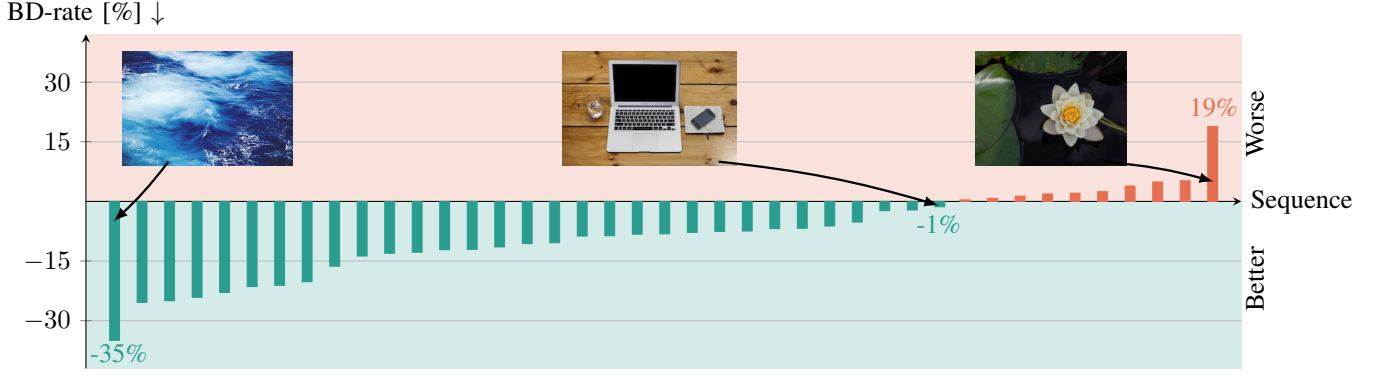
\begin{figure*}[t!]
    \centering
    \begin{tikzpicture}
      \begin{axis}[
        width=0.93\textwidth,
        height=6cm,
        enlargelimits=0.05,
        ybar,
        bar width=0.13cm,
        ybar=-0.1cm,                           
        xmin=0, xmax=42, minor x tick num=0, xtick style={draw=none}, xmajorticks=false, xlabel near ticks,
        ymin=-42, ymax=42, ytick distance={15}, minor y tick num=0,
        ymajorgrids,
        axis x line=middle,
        axis y line=middle,
        every axis x label/.style={at={(current axis.right of origin)},anchor=west},
        every axis y label/.style={at={(current axis.above origin)},anchor=south},
        xlabel={Sequence},
        ylabel={BD-rate [\%] $\downarrow$},
    ]
        \begin{pgfonlayer}{bg}
            \draw [cc_green!20, fill=cc_green!20] (axis cs:0,0) rectangle (axis cs:42,-42);
            \draw [cc_red!20, fill=cc_red!20] (axis cs:0,0) rectangle (axis cs:42,42);
            \node[rotate=90] at (axis cs: 42.5,19.5) {Worse};
            \node[rotate=90] at (axis cs: 42.5,-19.5) {Better};
        \end{pgfonlayer}

        \addplot [cc_red, fill=cc_red] table [col sep=comma] {data/clic20-bdrate-seq-positive.csv};
        \addplot [cc_green, fill=cc_green] table [col sep=comma] {data/clic20-bdrate-seq-negative.csv};

        \node [below right] at (axis cs:1,40) {\includegraphics[height=1.5cm]{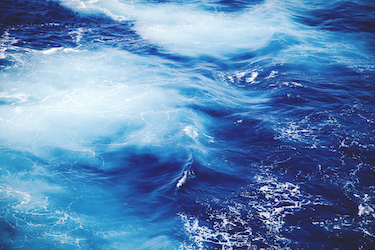}};
        \draw[->,-latex, line width=0.3mm] (axis cs:3, 10) to[bend left=5] (axis cs:1.0, -5);

        \node [below right] at (axis cs:17,40) {\includegraphics[height=1.5cm]{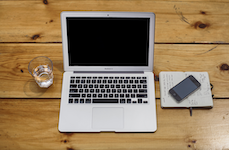}};
        \draw[->,-latex, line width=0.3mm] (axis cs:23, 10) to[bend left=5] (axis cs:31.0, -1);

        \node [below right] at (axis cs:32,40) {\includegraphics[height=1.5cm]{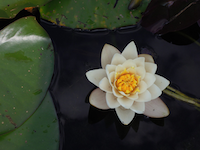}};
        \draw[->,-latex, line width=0.3mm] (axis cs:37, 10) to[bend left=5] (axis cs:41.0, 5);

        \node [below] at (axis cs:1.2,-33.5) {\textcolor{cc_green}{-35\%}};
        \node [below] at (axis cs:31,-1) {\textcolor{cc_green}{-1\%}};
        \node [above] at (axis cs:41,19) {\textcolor{cc_red}{19\%}};


    \end{axis}
    \end{tikzpicture}

    \caption{Sequence-wise BD-rate of the proposed system versus VVC (VTM 28.3)
    on the CLIC20 professional validation dataset. }
    \label{fig:sequence-wise-rate}
\end{figure*}

\begin{figure*}[t!]
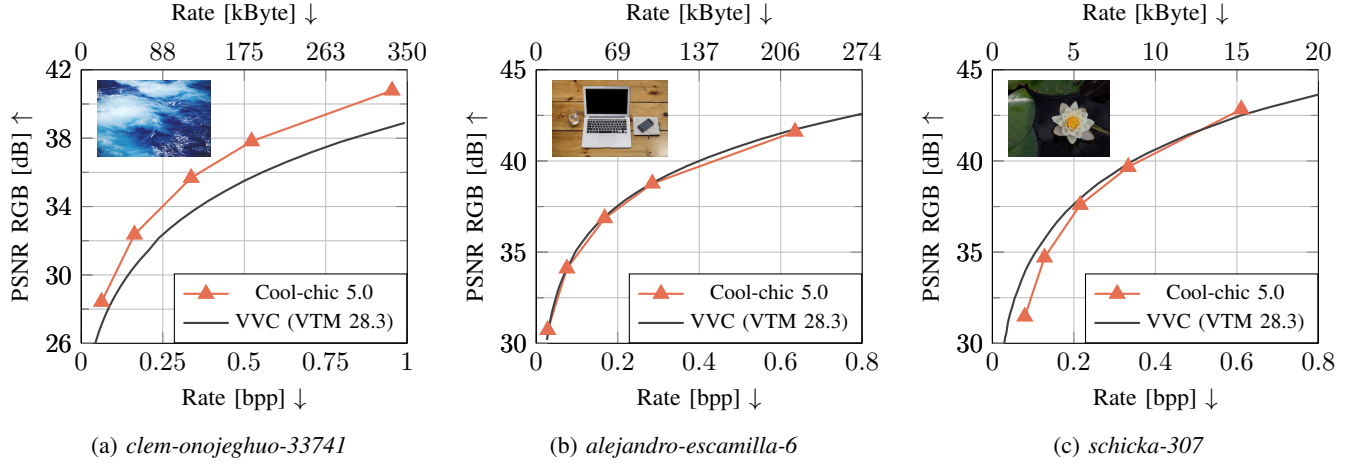

    \centering
        \begin{subfigure}{0.325\textwidth}
            \centering
            \begin{tikzpicture}
            \begin{axis}[
                grid= both,
                xlabel = {\small Rate [bpp] $\downarrow$},
                ylabel = {\small PSNR RGB [dB] $\uparrow$ } ,
                xmin = 0., xmax = 1,
                ymin = 26, ymax = 42,
                ylabel near ticks,
                xlabel near ticks,
                width=\linewidth,
                height=5.2cm,
                xtick distance={0.25},
                ylabel style={yshift=-2pt},
                xticklabel style={
                    /pgf/number format/fixed,
                    /pgf/number format/precision=2
                },
                ytick={26,30,34,38,42},
                minor y tick num=1,
                minor x tick num=0,
                legend style={at={(1.0,0.)}, anchor=south east},
                axis x line*=bottom,
            ]

            \addplot[thick, cc_red, mark=triangle*, mark size=3pt]
            coordinates {
                (0.953373, 40.786195)
                (0.52229, 37.818113)
                (0.336875, 35.678956)
                (0.163045, 32.363271)
                (0.062103, 28.427347)
            };
            \addlegendentry{\footnotesize Cool-chic 5.0}

            \addplot[thick, darkgray]
            coordinates {
                (0.992462225,	38.89817247)
                (0.887889194,	38.31480148)
                (0.793458104,	37.77019353)
                (0.706601992,	37.18212646)
                (0.630465888,	36.61807118)
                (0.560877404,	36.05270019)
                (0.498157051,	35.48658186)
                (0.441080014,	34.90217732)
                (0.391546474,	34.35472282)
                (0.345166552,	33.77990373)
                (0.30423821,	33.22015103)
                (0.269900412,	32.69406894)
                (0.23655277,	32.13530812)
                (0.206175595,	31.430002)
                (0.180062386,	30.88641321)
                (0.15674794,	30.33654761)
                (0.136584249,	29.80188216)
                (0.119119162,	29.28901111)
                (0.103325321,	28.77588542)
                (0.088075206,	28.2309016)
                (0.075998741,	27.73561479)
                (0.064840888,	27.22944378)
                (0.054884959,	26.72700936)
                (0.046394231,	26.21797878)
                (0.03907967,	25.74327742)
                (0.032858288,	25.29432902)
                (0.027358059,	24.86125818)
                (0.022475962,	24.39277898)
            };
        \addlegendentry{\footnotesize VVC (VTM 28.3)}

            \node [above right] at (rel axis cs:0.02,0.65) {\includegraphics[height=1cm]{figures/visual_comparison/clem-onojeghuo-33741-small.png}};
            \end{axis}
            \begin{axis}[
                axis x line*=top,
                xmin=0, xmax=350, 
                axis y line*=left,
                ymin = 26, ymax = 42,
                ytick={26,30,34,38,42},
                xtick={0, 87.5, 175, 262.5, 350},
                xticklabel style={
                    /pgf/number format/fixed,
                    /pgf/number format/precision=0
                },
                minor y tick num=1,
                width=\linewidth,
                height=5.2cm,
                xlabel = {\small Rate [kByte] $\downarrow$},
                xlabel near ticks,
            ]
            \end{axis}
        \end{tikzpicture}
        \caption{\textit{clem-onojeghuo-33741}}
        \label{fig:seq-wise-rd-clem}
    \end{subfigure}
    \begin{subfigure}{0.325\textwidth}
        \centering
        \begin{tikzpicture}
        \begin{axis}[
            grid= both,
            xlabel = {\small Rate [bpp] $\downarrow$},
            ylabel = {\small PSNR RGB [dB] $\uparrow$ } ,
            xmin = 0., xmax = 0.8,
            ymin = 30, ymax = 45,
            ytick distance={5},
            ylabel near ticks,
            xlabel near ticks,
            width=\linewidth,
            height=5.2cm,
            xtick distance={0.2},
            ylabel style={yshift=-2pt},
            xticklabel style={
                /pgf/number format/fixed,
                /pgf/number format/precision=2
            },
            minor y tick num=1,
            minor x tick num=0,
            legend style={at={(1.0,0.)}, anchor=south east},
            axis x line*=bottom,
        ]

        \addplot[thick, cc_red, mark=triangle*, mark size=3pt]
        coordinates {
            (0.635317, 41.607776)
            (0.284768, 38.764199)
            (0.167997, 36.862403)
            (0.075243, 34.10573)
            (0.028285, 30.733188)
        };
        \addlegendentry{\footnotesize Cool-chic 5.0}

        \addplot[thick, darkgray]
        coordinates {
            (0.931411618,	43.19565132)
            (0.813512299,	42.65981708)
            (0.71073329,	42.13778324)
            (0.627027516,	41.6883669)
            (0.551948166,	41.20364461)
            (0.482965949,	40.69566)
            (0.422464292,	40.20545014)
            (0.370600728,	39.73016163)
            (0.324052465,	39.24571164)
            (0.28293911,	38.76688897)
            (0.248188364,	38.29959245)
            (0.217323446,	37.83745573)
            (0.190819875,	37.37432903)
            (0.166092933,	36.90854486)
            (0.14543561,	36.44380971)
            (0.1271092,	35.97457857)
            (0.111055358,	35.49913427)
            (0.09693568,	35.05610404)
            (0.08428048,	34.48312792)
            (0.073696555,	34.02743676)
            (0.064530433,	33.5599986)
            (0.056493302,	33.11205456)
            (0.050550784,	32.68840715)
            (0.04434571,	32.19796074)
            (0.038656997,	31.71742053)
            (0.03435983,	31.24919531)
            (0.03017352,	30.76170332)
            (0.026188504,	30.19099941)
        };
        \addlegendentry{\footnotesize VVC (VTM 28.3)}
        \node [above right] at (rel axis cs:0.02,0.65) {\includegraphics[height=1cm]{figures/visual_comparison/alejandro-escamilla-6-small.png}};
        \end{axis}
        \begin{axis}[
            axis x line*=top,
            xmin=0, xmax=274.2272, 
            axis y line*=left,
            ymin = 30, ymax = 45,
            ytick distance={5},
            xtick={0, 68.5, 137, 205.5, 274},
            xticklabel style={
                /pgf/number format/fixed,
                /pgf/number format/precision=0
            },
            minor y tick num=1,
            width=\linewidth,
            height=5.2cm,
            xlabel = {\small Rate [kByte] $\downarrow$},
            xlabel near ticks,
        ]
        \end{axis}
    \end{tikzpicture}
    \caption{\textit{alejandro-escamilla-6}}
    \label{fig:seq-wise-rd-alejandro}
    \end{subfigure}
    \begin{subfigure}{0.325\textwidth}
        \centering
        \begin{tikzpicture}
        \begin{axis}[
            grid= both,
            xlabel = {\small Rate [bpp] $\downarrow$},
            ylabel = {\small PSNR RGB [dB] $\uparrow$ } ,
            xmin = 0., xmax = 0.8,
            ymin = 30, ymax = 45,
            ytick distance={5},
            ylabel near ticks,
            xlabel near ticks,
            width=\linewidth,
            height=5.2cm,
            xtick distance={0.2},
            ylabel style={yshift=-2pt},
            xticklabel style={
                /pgf/number format/fixed,
                /pgf/number format/precision=2
            },
            minor y tick num=1,
            minor x tick num=0,
            legend style={at={(1.0,0.)}, anchor=south east},
            axis x line*=bottom,
        ]

        \addplot[thick, cc_red, mark=triangle*, mark size=3pt]
        coordinates {
            (0.610824, 42.832588)
            (0.333397, 39.674476)
            (0.215546, 37.593929)
            (0.127883, 34.708124)
            (0.079425, 31.468103)
        };
        \addlegendentry{\footnotesize Cool-chic 5.0}

        \addplot[thick, darkgray]
        coordinates {
            (1.029378255,	44.82311431)
            (0.882364909,	44.10471695)
            (0.779866536,	43.52483863)
            (0.681844076,	42.95881768)
            (0.610717773,	42.50722464)
            (0.543986003,	41.96172221)
            (0.480957031,	41.44752885)
            (0.429117839,	40.92831167)
            (0.379557292,	40.41818779)
            (0.332967122,	39.86344348)
            (0.294677734,	39.28366707)
            (0.259643555,	38.76571848)
            (0.231974284,	38.2642545)
            (0.203653971,	37.70393296)
            (0.180460612,	37.21132944)
            (0.157999674,	36.64281431)
            (0.14050293,	36.12876431)
            (0.121704102,	35.51937215)
            (0.106770833,	35.03645418)
            (0.094156901,	34.56494774)
            (0.08190918,	34.03530484)
            (0.072021484,	33.50031342)
            (0.061889648,	32.88392911)
            (0.054280599,	32.45852136)
            (0.048502604,	31.9858383)
            (0.043375651,	31.62882966)
            (0.037760417,	31.11267328)
            (0.034179688,	30.56305812)
            (0.030965169,	30.29111998)
            (0.02750651,	29.76187815)
            (0.024332682,	29.43637301)
            (0.021402995,	28.77022652)
            (0.018636068,	28.33049797)
            (0.016398112,	27.84248225)
        };
        \addlegendentry{\footnotesize VVC (VTM 28.3)}
        \node [above right] at (rel axis cs:0.02,0.65) {\includegraphics[height=1cm]{figures/visual_comparison/schicka-307-small.png}};
        \end{axis}
        \begin{axis}[
            axis x line*=top,
            xmin=0, xmax=19.6, 
            axis y line*=left,
            ymin = 30, ymax = 45,
            ytick distance={5},
            xtick={0, 4.9, 9.8, 14.7, 19.6},
            xticklabel style={
                /pgf/number format/fixed,
                /pgf/number format/precision=0
            },
            minor y tick num=1,
            width=\linewidth,
            height=5.2cm,
            xlabel = {\small Rate [kByte] $\downarrow$},
            xlabel near ticks,
        ]
        \end{axis}
    \end{tikzpicture}
    \caption{\textit{schicka-307}}
    \label{fig:seq-wise-rd-schicka}
\end{subfigure}
\caption{Rate-distortion graph on 3 images from the CLIC20 professional validation dataset.}
\label{fig:seq-wise-rd}
\end{figure*}

This section studies the performance of Cool-chic 5.0 for the different
sequences of the CLIC20 dataset. Figure \ref{fig:sequence-wise-rate} presents
the sequence-wise BD-rates obtained against VVC. The proposed system outperforms
VVC for most of the sequences, obtaining up to 35\% BD-rate reduction for the
best case. In order to better understand the performance of Cool-chic 5.0, we
provide rate-distortion plots on 3 sequences, visible in Fig.
\ref{fig:seq-wise-rd}.
\newline

Cool-chic appears to be more performant than conventional codecs for images
featuring rich, non-directional textures such as the one presented in Fig.
\ref{fig:seq-wise-rd-clem}. For this image, Cool-chic consistently achieves
better quality than VVC for the entire rate range evaluated. Cool-chic and VVC
offer similar compression performance for content presenting many directional
edges, like the one presented in Fig. \ref{fig:seq-wise-rd-alejandro}. For
Cool-chic, the worst image of the CLIC20 dataset comparatively to VVC is shown
in Fig. \ref{fig:seq-wise-rd-schicka}. For this image, Cool-chic requires 19\%
more rate than VVC to reach similar quality. This image turns out to be an image
of small resolution ($512 \times 384$) with limited high frequencies, resulting
in a low file size of a few kBytes. At these lower rates, the share of the neural
networks parameter starts to be comparable to the latent grids, causing a
substantial rate overhead that saturates the compression performance.

\subsection{Rate of the neural network parameters}
\label{sec:rate-nn-experiments}

\begin{figure}[bt]
    \centering
    \begin{tikzpicture}
    \centering
    \begin{semilogxaxis}[
        grid= major,
        xlabel = {\small Compressed file size [kByte]},
        ylabel = {\small Neural network parameters size [kByte]},
        xmin = 1, xmax = 1000,
        ymin = 0, ymax = 3,
        ytick distance={0.5},
        ylabel near ticks,
        xlabel near ticks,
        width=\linewidth,
        height=6cm,
        x tick label style={/pgf/number format/1000 sep=\,},
        log base 10 number format code/.code={%
            $\pgfmathparse{10^(#1)}\pgfmathprintnumber{\pgfmathresult}$%
        },%
        ylabel style={yshift=-2pt},
        legend style={at={(0.95,0.4)}, anchor= east, },
        legend cell align={left},
        ymajorgrids=true,
        yminorgrids=true,
    ]
        \addplot [cc_red, mark=*, only marks, mark size=1.25pt] table [col sep=space, x=rate_kbytes, y=nn_kbytes] {data/rate_nn_clic20.csv};
        \addlegendentry{\footnotesize{CLIC20 pro validation}}

        \addplot [cc_blue, mark=square*, only marks, mark size=1.25pt] table [col sep=space, x=rate_kbytes, y=nn_kbytes] {data/rate_nn_kodak.csv};
        \addlegendentry{\footnotesize{Kodak}}

    \end{semilogxaxis}
    \end{tikzpicture}
    \caption{Size of the compressed neural network parameters as a function of
    the file size for each image. HOP decoder configuration (1\,900
    parameters).}
    \label{fig:rate-nn}
\end{figure}
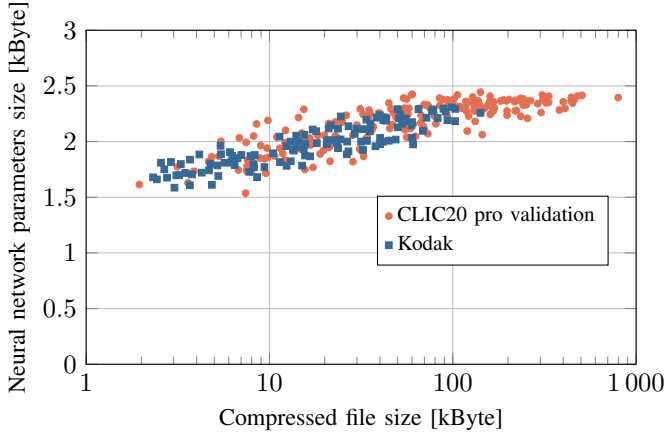

Since they are adapted to each individual image, the parameters of the neural
networks composing the decoder are transmitted alongside the latent
representation. Figure \ref{fig:rate-nn} represents the rate of the neural
network parameters for all images of the Kodak and CLIC20 datasets, compressed
under the 5 rate constraints $\lambda$ listed in Table
\ref{table:encoder-parameters}. While the compressed file size varies from a few
kBytes to almost 1\,000 kBytes, the rate of the neural networks remains
approximately constant ranging from 1.5 to 2.5 kBytes.
\newline

When the compressed file size is big enough, the overhead due to neural network
has a marginal impact on the overall compression performance. However, for
aggressive rate constraint and smaller images, the compressed file size is
smaller than 10 kBytes. Here, the rate overhead caused by neural network
parameters has a noticeable impact on the performance as it represent 20\% or
more of the overall rate. This might explain the relatively worse compression
efficiency of Cool-chic on the Kodak dataset, which features smaller images than
the CLIC20 dataset as presented in Fig. \ref{fig:seq-wise-rd-schicka}.

\section{Ablation study}

This section shows the performance gain brought by Cool-chic 5.0 contributions.
It first reconsiders the encoding parameters presented in Table
\ref{table:encoder-parameters} and then the proposed decoder changes. Table
\ref{table:ablation results} summarizes the performance gain brought by each of
the proposed contribution on the CLIC20 dataset, using the HOP decoder (see
Table \ref{table:decoder-parameters}) configuration and 100\,000 encoding iterations.

As a reference, Table \ref{table:ablation results} also presents the performance
of the previous Cool-chic 4.2 release. Cool-chic 5.0 offers around 10\% rate
reduction compared to the previous Cool-chic 4.2 release. This improvement is
brought by contributions to both the encoder and the decoder, with similar gains
on both sides of the codec.

\subsection{Encoder changes}

Cool-chic 5.0 presents substantial improvements to the encoding process. We
propose to evaluate the gains brought by two contributions: the usage of a
second-order optimizer and a revised differentiable proxy for the quantization
function $Q_{train}$. Since these changes affect only the encoding process, the
decoder complexity remains identical for both ablations.

The \textit{No SOAP} configuration replaces the SOAP optimizer by the
first-order Adam optimizer for the neural networks. This
results in worse convergence and $+2.8\%$ BD-rate increase.

The \textit{C3 $Q_{train}$} configuration uses the differentiable quantization
proxy proposed in C3 \cite{c3-kim}. The random noise $\mathbf{n}_{\sigma}$
follows a Kumaraswamy distribution whose parameter is scheduled linearly from
$2.0$ to $1.0$ and the softround temperature $T$ goes from $0.3$ to $0.1$ during
the main training stage. This \textit{C3 $Q_{train}$} configuration results in
worse convergence, with a BD-rate increase of $+2.7\%$.

\begin{table}[t]
    \caption{Ablation of the proposed contributions evaluated relative to the
    proposed Cool-chic 5.0 with the HOP decoder configuration. All
    configurations are tested on the CLIC20 pro validation dataset with 100\,000
    encoding iterations.}
    \centering
    \small
    \begin{tblr}{
        colspec={Q[l] Q[c] Q[c]},
        row{1} = {font=\bfseries,darkgray!5, belowsep=0.25mm},
    }
        System          & BD-rate $\downarrow$ & Decoder MAC/pixel   \\
        \cmidrule{1-Z}
        Cool-chic 5.0 (HOP) & 0.0\%     & 1\,991   \\
        Cool-chic 4.2   & 10.4\%    & 1\,432       \\
        \cmidrule{1-Z}
        No SOAP         & +2.8\%    & 1\,991        \\
        C3 $Q_{train}$  & +2.7\%    & 1\,991        \\
        \cmidrule{1-Z}
        No IFCE         & +3.8\%    & 1\,955        \\
        No hyperlatent  & +0.5\%    & 1\,978        \\
        No stabilizer   & +0.4\%    & 1\,914        \\
    \end{tblr}
    \label{table:ablation results}
\end{table}

\subsection{Decoder architecture}

This paper introduces three main changes to the decoder architecture. IFCEs are
introduced to leverage inter feature redundancies, hyperlatent grids are added
to improve the entropy modeling and a stabilizer branch is added to the ARM and
synthesis network, helping the convergence.

The \textit{No IFCE} configuration removes the inter feature context
$\mathbf{f}$ from the ARM input (see Fig. \ref{fig:IFCE}). To maintain similar
decoding complexity, these inter feature contexts are replaced by additional
spatial contexts $\mathbf{s}$. The results show that removing the IFCE module
significantly reduces the compression performances with a BD-rate increase of
$+3.8\%$.

The \textit{No hyperlatent} configuration removes the hyperlatent grids
$\qlatent_h$, resulting in an increased BD-rate of $+0.5\%$.

The \textit{No stabilizer} configuration removes the stabilizer layers from the
ARM and the synthesis \textit{i.e.,} leaving only the trunk branch. Removing
this results in a BD-rate increase of $+0.4\%$.

\section{Future work}

\subsection{Sequence-wise parameters adaptation}

This paper shows that the hyperparameters related to training have a
significant impact on the performance \textit{e.g.,} refining the quantization
proxy or the optimizer yields important rate savings. Yet, this work still uses
a single set of hyperparameters for all images even though it is likely that
the hyperparameters should be adapted to the content. A relevant future work
is to automatically derive the optimal hyperparameters from the content.

The initialization of the neural networks and the latent grids is also common to
all images in this work. Earlier work \cite{hypercool-pep} showed that
meta-learned initialization allows for faster convergence for the first few
hundreds training iterations at the expense of the asymptotic performance. This
implies that having content based initialization would be beneficial and lead to
faster encodings.
\newline

Finally, the rate dedicated to the neural networks parameters have been shown to
be too important for some rate targets. One explanation for this is that the
same decoder architecture is used for all images and rate targets. One
workaround would be to automatically switch to smaller architectures, based on
the rate constraint $\lambda$ and the image properties, allowing to efficiently
address these lower rates.

\subsection{Better neural network compression}

This work follows the encoding process proposed in the initial Cool-chic paper,
where the neural network rate is ignored during the training stage. While
justifiable for most of the use-cases, this causes suboptimal performance at
lower rates. Introducing the neural network rate in the objective function of
the optimization process could help to improve the compression performance. This
could be done through the addition of sparsity-based or entropy-based
constraints of the weights and biases of the decoder.
\newline

Quantization-aware Training (QAT) is also a promising idea to reduce the rate of
neural networks, by preparing them to be encoded into smaller amount of bits
\textit{e.g.} 4-bit parameters. QAT has been shown to be successful on large
models such as Large Language Model (LLM) \cite{qat-llm-bondarenko} but has not
been explored yet for the smaller neural networks found in overfitted codecs.

\section{Conclusion}

This paper presents Cool-chic 5.0, the latest version in the Cool-chic series of
codecs, refining both the encoder and decoder. New decoder tools are introduced
such as the Inter-Feature Context Extractor or hyperlatent grids yielding a
richer entropy model and better compression performance. Significant effort is
made to speed up the encoding process either through the usage of a second-order
optimizer, a refining of the proxy function for the quantization relaxation and
the introduction of stabilizer linear layers at the decoder.

The resulting Cool-chic 5.0 improves the performance of overfitted codecs, with
more than 7\% rate savings while keeping a low decoding complexity of 2\,000
multiplications per decoded pixel. It is the first overfitted codec to be
competitive with modern autoencoders such as MLIC++ and to offer more than 11\%
rate savings compared to the state-of-the-art conventional codec H.266/VVC.
Finally, Cool-chic 5.0 also divides by 10 the number of encoding iterations
compared to other overfitted codecs for identical performance. In summary,
Cool-chic 5.0 represents a substantial improvement for overfitted codecs,
featuring enhanced compression performance and faster encoding, thereby
demonstrating the relevance and potential of this new compression paradigm.


\bibliographystyle{IEEEtran}
\bibliography{
    bibliography/ai,
    bibliography/autoencoder,
    bibliography/conventional,
    bibliography/coolchic,
    bibliography/dataset_metrics,
    bibliography/misc,
    bibliography/overfitted
}


\section{Biography Section}




\begin{IEEEbiography}[{\includegraphics[width=1in,height=1.25in,clip,keepaspectratio]{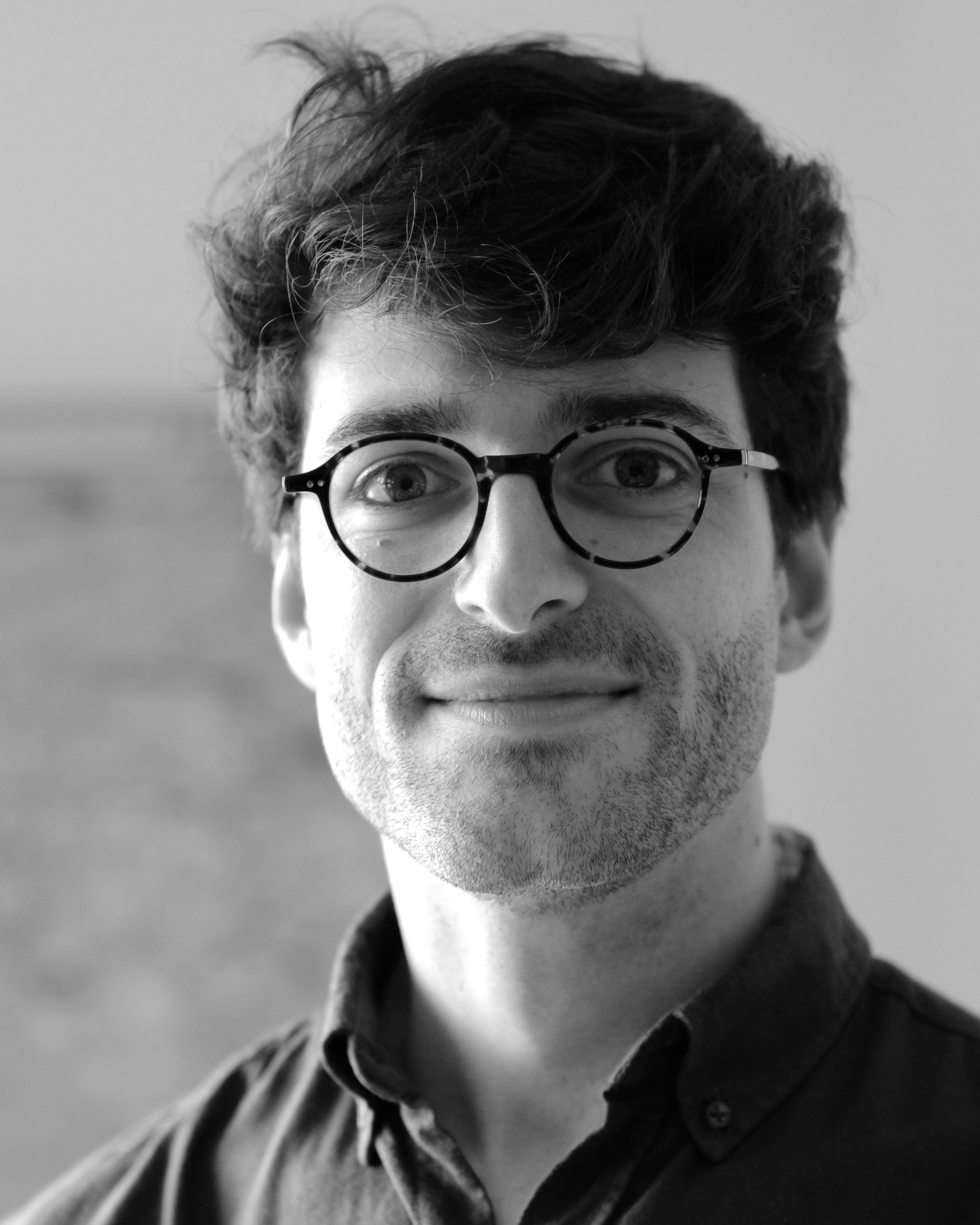}}]{Théo
Ladune} received his M.S. degree in Computer Engineering from Rennes University
in 2018. He then obtained a Ph.D. degree in signal processing from the same
institution in 2021, working on deep learning-based video codecs. Since joining
Orange in 2018, he has been investigating various applications of deep learning
in image and video coding, including work on autoencoders, overfitted codecs.
His current research focuses on creating innovative video coding algorithms that
rely on lightweight neural networks.
\end{IEEEbiography}

\begin{IEEEbiography}[{\includegraphics[width=1in,height=1.25in,clip,keepaspectratio]{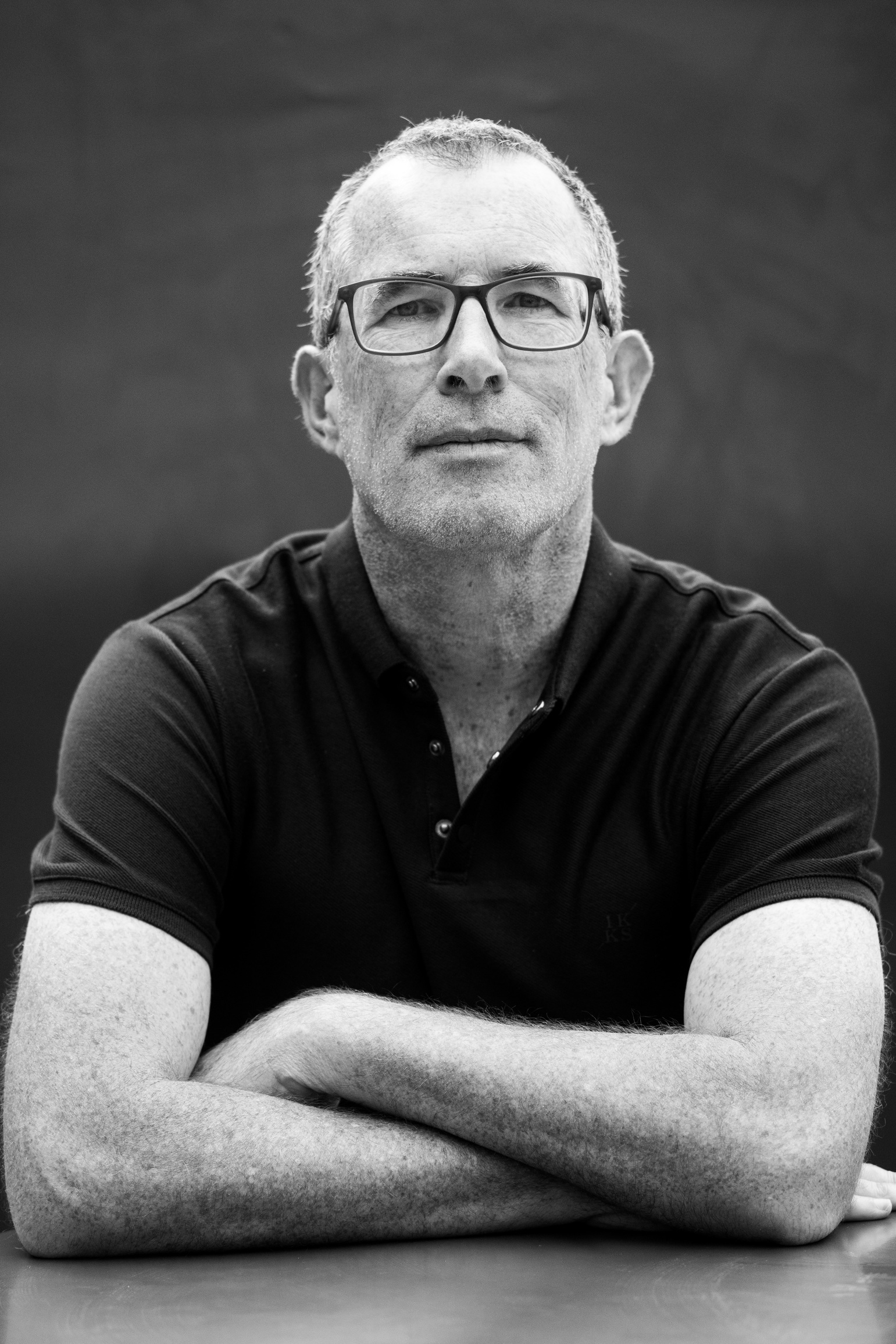}}]{Pierrick
Philippe} received the Ph.D. degree in signal processing from the University of
Paris Sud, Orsay, in 1995. Before joining Orange Labs, he spent two years at
Innovason developing digital mixing desks, and he also spent two years at
Envivio, where his activities focused on audio signal processing. At Orange, he
was involved in the development of audio coding technologies, such as HE-AAC. He
is currently a Senior Video Coding Specialist with Orange Research, developing
video algorithms, especially related to video coding standards (MPEG). His main
interest is signal processing, especially audiovisual signal representation and
compression.
\end{IEEEbiography}

\begin{IEEEbiography}[{\includegraphics[width=1in,height=1.25in,clip,keepaspectratio]{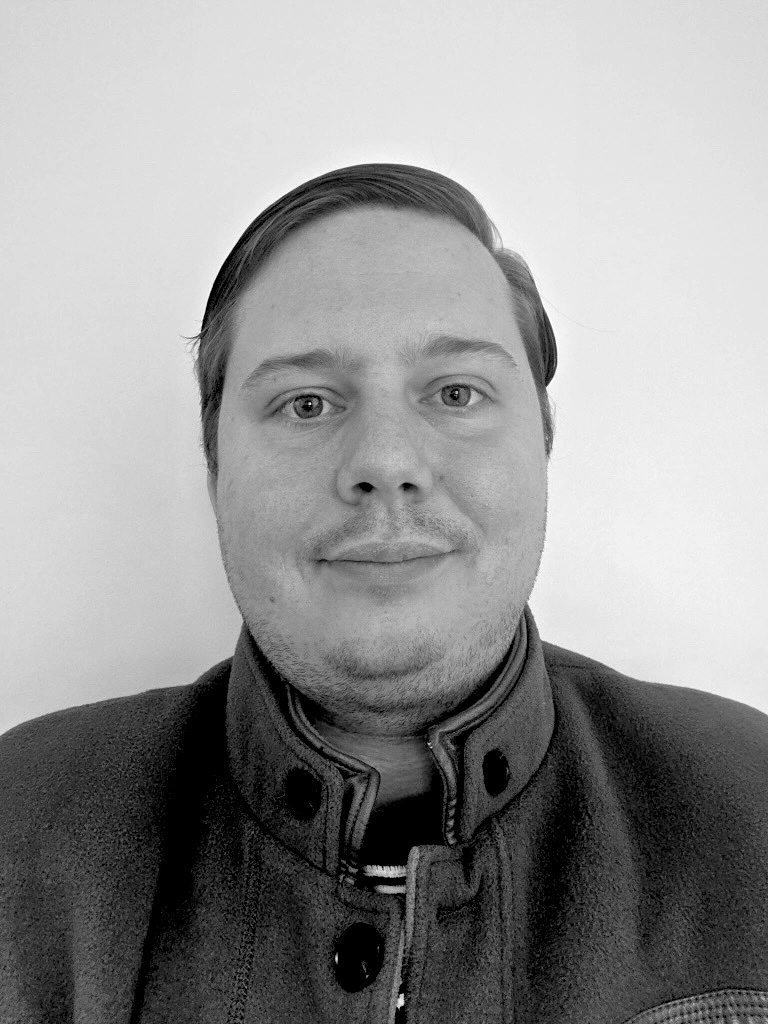}}]{Pierre
Jaffuer} received the M.Sc degree in computer science from the University of La
Réunion in 2023. He then worked for two years in the industry as a software
engineer. In 2025, he started a Ph.D. in signal processing at the University of
Rennes in collaboration with Orange. His main research interest is
learning-based image and video compression.
\end{IEEEbiography}

\begin{IEEEbiography}[{\includegraphics[width=1in,height=1.25in,clip,keepaspectratio]{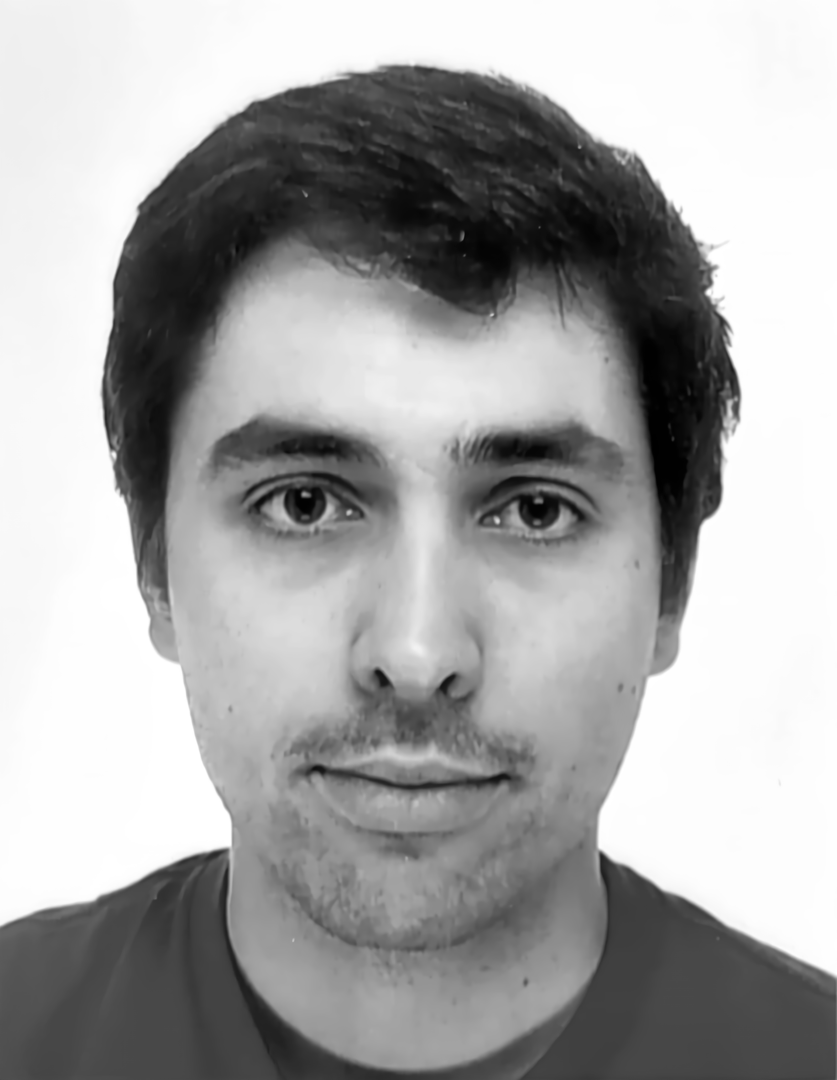}}]{Théophile
Blard} received his M.S. degree in Computer Engineering from the University of
Strasbourg in 2019, where he also obtained a second master’s degree focused on
computer vision. He then worked for several years in industry as a Machine
Learning Engineer, focusing on deep learning-based image and video processing.
He is currently pursuing a Ph.D. in signal processing at the University of
Rennes, in collaboration with Orange, starting in 2023. His research focuses on
low-complexity learned image and video compression.
\end{IEEEbiography}

\begin{IEEEbiography}[{\includegraphics[width=1in,height=1.25in,clip,keepaspectratio]{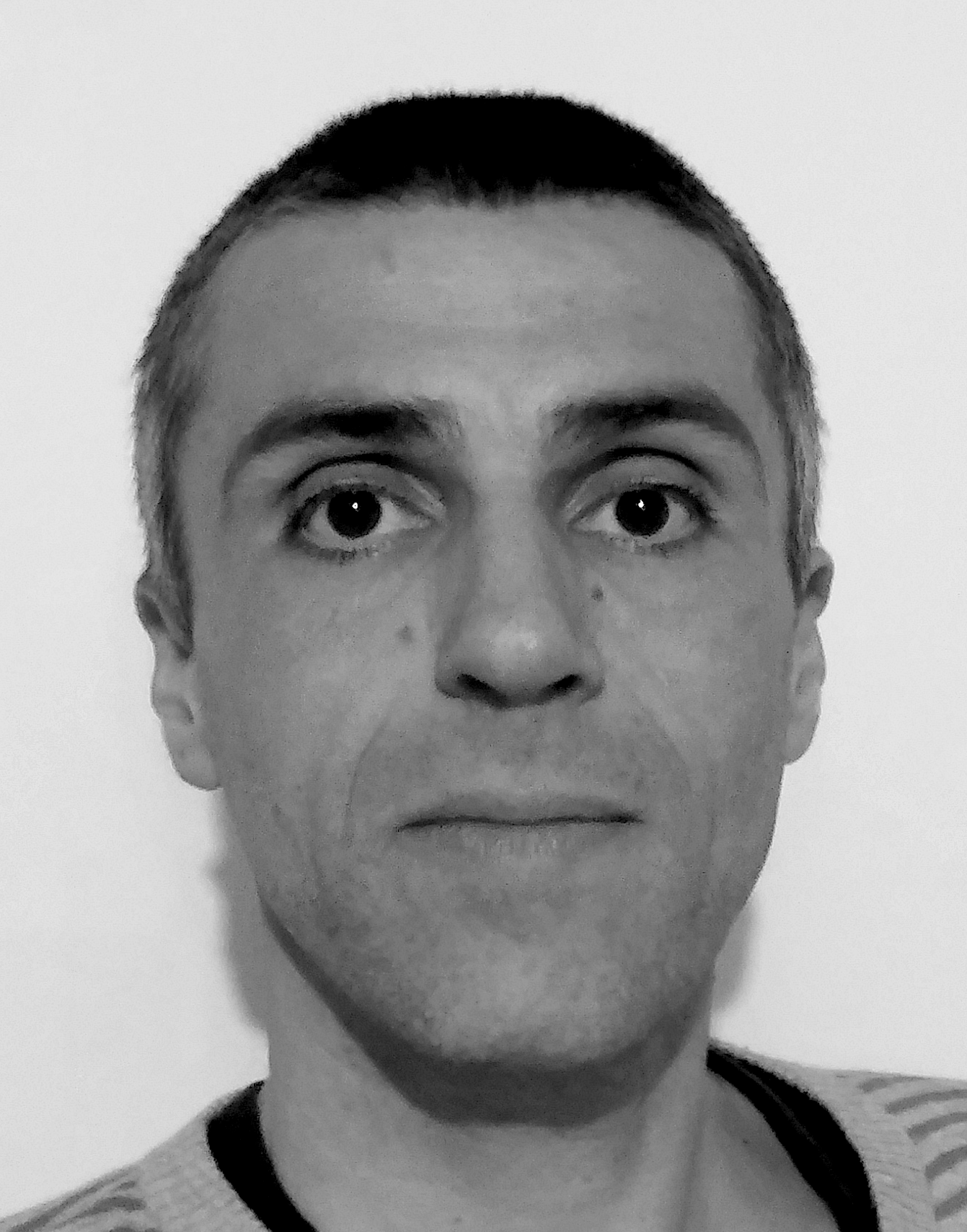}}]{Sylvain
Kervadec} received the M.S. Degree in computer science and image processing in
2001 from the University of Rennes 1, France. In 2003, he joined France Telecom
/ Orange, where he actively participates in the commercial deployment of new
audiovisual technologies. He has successfully contributed to several standards,
notably the scalable extensions of MPEG-4 AVC, and MPEG DASH. His research
interests include image and video compression and communication.
\end{IEEEbiography}

\begin{IEEEbiography}[{\includegraphics[width=1in,height=1.25in,clip,keepaspectratio]{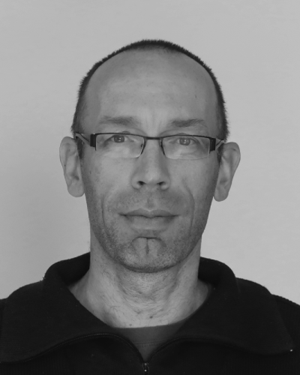}}]{Félix Henry} received the Dipl.-Ing.
(M.Sc.) degree in telecommunications from Telecom ParisSud, Paris, France, in
1993, and the Ph.D. degree from Telecom ParisTech, Paris, in 1998 in the domain
of wavelet image coding. He started his career in 1995 with the Canon Research
Center, France, where he worked on still image compression and video coding. He
is currently working as a Research Engineer with Orange Labs and b<>com National
Research Institute, Rennes, France. He participated actively in JPEG2000, HEVC,
and VVC standardization and is a co-inventor of more than 150 patents in the
domain of signal processing. His current research interests include immersive
video coding and video compression using neural networks
\end{IEEEbiography}

\begin{IEEEbiography}[{\includegraphics[width=1in,height=1.25in,clip,keepaspectratio]{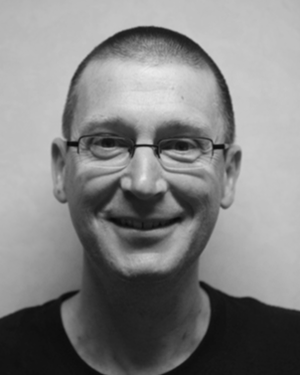}}]{Gordon Clare} received the M.Sc. degree from the
University of Auckland, Auckland, New Zealand, in 1984.,From 1985 to 1989, he
was a Development Engineer with Rakon Computers, Sydney, Australia. from 1989 to
1991, he was with Software Development International, heading a team of
developers creating a network management system for fault tolerant environments.
from 1991 to 1997, he was with CISRA, a Canon Research Center, Sydney, focusing
on real-time hardware and software image processing solutions. After moving to
France in 1997, he was with several international companies, developing document
management systems and a search engine. As a Consultant from 2005 to 2010, he
was developing H.264 and scalable video coding (SVC) solutions at Canon and
France Telecom-Orange, Rennes, France. Since 2010, he has been with France
Telecom-Orange, continuing his research related to HEVC and VVC standardization
and development. His main research interest is the usage of neural networks for
video compression.
\end{IEEEbiography}

\vfill

\end{document}